\documentstyle[prc,aps,preprint,psfig]{revtex}

\draft

\begin{document}
\title{Hadronic observables from SIS to SPS energies --  anything strange
with strangeness ?}

\author{H. Weber$^1$, E.~L.~Bratkovskaya$^1$\thanks{Supported by DFG},
W. Cassing$^2$ and H.~St\"ocker$^{1,3}$\\[5mm]
{$^1$ \normalsize Institut f\"{u}r Theoretische Physik,
    Universit\"{a}t Frankfurt}\\
    {\normalsize 60054 Frankfurt, Germany}\\[3mm]
{$^2$ \normalsize Institut f\"{u}r Theoretische
    Physik, Universit\"{a}t Giessen}\\
    {\normalsize 35392 Giessen, Germany} \\[3mm]
{$^3$ \normalsize SUBATECH, Laboratoire de Physique Sabatomique
    et des Technologies Associ\'ees}\\
    {\normalsize University of Nantes - IN2P3/CNRS - Ecole des
     Mines de Nantes}\\
    {\normalsize 4 rue Alfred Kastler, F-44072 Nantes, Cedex 03, France}}
\maketitle

\begin{abstract}
We calculate $p, \pi^\pm, K^\pm$ and $\Lambda$(+$\Sigma^0$)  rapidity
distributions and compare to experimental data from SIS to SPS
energies within the UrQMD and HSD transport approaches that are both
based on string, quark, diquark ($q, \bar{q}, qq, \bar{q}\bar{q}$) and
hadronic degrees of freedom. The two transport models do not include any
explicit phase
transition to a quark-gluon plasma (QGP). It is found that both
approaches agree rather well with each other and with the
experimental rapidity distributions for protons, $\Lambda$'s, $\pi^\pm$
and $K^\pm$.  Inspite of this apparent agreement both transport models
fail to reproduce the maximum in the excitation function for the ratio
$K^+/\pi^+$ found experimentally between 11 and 40 A$\cdot$GeV. A
comparison to the various experimental data shows that this 'failure' is
dominantly due to an insufficient description of pion rapidity
distributions rather than missing 'strangeness'.  The modest differences in
the transport model results -- on the other hand -- can be attributed
to different implementations of string formation and fragmentation,
that are not sufficiently controlled by experimental data for the
'elementary' reactions in vacuum.
\end{abstract}

\noindent PACS: 24.10.-i; 25.75.-q; 11.30.Rd; 13.60.-r

\noindent Keywords: Nuclear-reaction models and methods;
Relativistic heavy-ion collisions; Chiral symmetries; Meson
production

\newpage
\narrowtext
\section{Introduction}

The ultimate goal of relativistic nucleus-nucleus collisions is to
reanalyse the early 'big-bang' under laboratory conditions and to find
the 'smoking gun' for a phase transition from the expected initial
quark-gluon plasma (QGP) to a phase characterized by an interacting
hadron gas \cite{Harris,Shuryak93,QM99}. Though evidence for a 'new
phase of hadronic matter' at the SPS has been claimed \cite{Heinz}, a
direct proof -- according to the understanding of the authors -- is
still lacking \cite{Bass:1998vz,QM2001}.  Furthermore, nucleus-nucleus
collisions with initial energies per nucleon of $\approx$
21.3~A$\cdot$TeV ($\sqrt{s}$ = 200~GeV) are available now at the
Relativistic-Heavy-Ion-Collider (RHIC) in Brookhaven, an order of
magnitude higher than at SPS energies ($\sqrt{s} \approx$ 17-19 GeV).
In central collisions of Au~+~Au nuclei here energy densities above 4
GeV/fm$^3$ are expected \cite{QM2001}.  These estimates are based on
the Bjorken prescription \cite{Bjorken} employing a formation time of
$\tau$ = 1 fm/c. The latter quantity is uncertain by at least a factor
of 2 which implies a corresponding uncertainty in the energy density.
Nevertheless, energy densities of a few GeV/fm$^3$ suggest that the
critical energy density for a QGP phase should be overcome in
considerable space-time volumes at RHIC, where the relevant degrees of
freedom are partons (quarks and gluons).  Parton cascade calculations
\cite{Geiger3,Geiger1,Wang} are expected to provide suitable
descriptions in the early phase \cite{Baier:1998yf,Ruppert:2001yr} of
these collisions whereas hadrons should only be formed (by
'condensation') at a later stage which might be a couple of fm/c from
the initial contact of the heavy ions.  In fact, hybrid models like
VNI+UrQMD \cite{Bass:1999pv}, VNI+HSD \cite{VNIHSD} or the AMPT
approach \cite{CMKO} also allow for a reasonable description of the
'soft' hadronic observables so far, which -- due to the high
interaction rate -- are found to be close to the hydrodynamic limit
\cite{hydro}. On the other hand, once the local equilibrium limit is
reached in the reaction, any conclusion on the dynamics in the early
nonequilibrium phase and its dynamical degrees of freedom becomes
highly model dependent.

Moreover, the question of chiral symmetry restoration at high baryon
density and/or high temperature is of fundamental interest, too
\cite{Harris,Shuryak93}. Whereas lattice QCD calculations at zero
baryon chemical potential indicate that a restoration of chiral
symmetry goes along with the deconfinement phase transition at a
critical temperature $T_c$, the situation is less clear at finite
baryon density where QCD sum rule studies show a linear decrease of the
scalar quark condensate $\langle\bar q q\rangle$ -- which is
nonvanishing in the vacuum due to a spontaneous breaking of chiral
symmetry -- with baryon density $\rho_B$ towards a chiral symmetric
phase characterized by $\langle\bar{q} q\rangle$ = 0. This decrease of
the scalar condensate is expected to lead to a change of the hadron
properties with density and temperature, i.e.\ in a chirally restored
phase the hadrons might become approximately massless as suggested in
Ref. \cite{Brown}.  However, chiral symmetry restoration only implies
that vector and axial vector currents should become equal
\cite{Kochr,Zahed}. Thus vector and axial vector excitations of the QCD
vacuum must have the same spectral functions in the chiral limit.  As
demonstrated in Refs. \cite{Toneev98,HSD_exf} such a restoration of
chiral symmetry in central nucleus-nucleus collisions should  ---
driven by the baryon density --- occur at bombarding energies of 5--10
A$\cdot$GeV.  In Ref. \cite{GazdRoe95} it has been argued,
furthermore, that such 'phase transitions' should be seen in a much
lower strangeness to entropy ratio. It has been also suggested
\cite{HSD_exf} that especially the $K^+/\pi^+$ might give an indication
for a chirally restored phase.

The fact that the $K^+/\pi^+$ ratio is found experimentally to be
higher at top AGS energies of 11 A$\cdot$GeV than at 160 A$\cdot$GeV
has raised speculations about the appearance of 'new physics' at
energies between AGS and top SPS.  To shed some light on this issue,
the NA49 Collaboration has started an energy scan at the SPS. First
results have become available now at 40 and 80 A$\cdot$GeV
\cite{NA49_new,NA49_Lam,Misch02} and further studies are foreseen at 30
and 20 A$\cdot$GeV \cite{SPS20}. Since this topic is of current
interest we will restrict our investigations to the AGS and SPS energy
range in this paper.

 From the theoretical side the various hadron spectra are
conventionally calculated with nonequilibrium kinetic transport theory
(cf. \cite{CBRep98,Stoecker,Bertsch,Cass90,Koreview,WangSorge}).
However, the calculated kaon to pion ratio from central nucleus-nucleus
collisions turns out to vary by factors as large as 2 if different
transport approaches are applied
\cite{HSD_exf,Jgeiss,WBS_K02,WangSorge}. Thus a unique interpretation
of the data is questionable so far. On the other hand, statistical
models \cite{StatMod} show a maximum of $K^+/\pi^+$ ratio at $\sim 30$
A$\cdot$GeV since the relative strangeness content of baryons is
highest at low bombarding energies. It decreases with higher energies
due to an increase of temperature and a decrease of the baryon chemical
potential. However, an analysis within the UrQMD transport model
suggests that chemical and thermal equilibria are achieved only briefly
in a small central overlap region of heavy-ion collision due to a very
fast expansion of the hadronic fireball \cite{Bravina}.  Moreover, the
analysis of Ref. \cite{Brat_Thermo} within the HSD transport approach
indicates that the equilibration time for strangeness at all bombarding
energies is larger ($\geq$ 40 fm/c) than the reaction time of
nucleus-nucleus collisions.  Thus the statistical model fits to the
data have to be considered with some caution since they are not
understood microscopically.

In this work we concentrate on hadronic rapidity distributions of
protons, kaons, antikaons and hyperons and their yields and ratios from
$Au + Au$ (or $Pb + Pb$) collisions from SIS to SPS energies. The aim
of our study is twofold: first, to find out the systematic differences
between two currently used transport approaches (denotes as UrQMD
\cite{UrQMD1,UrQMD2} and HSD \cite{CBRep98,Ehehalt}) and second, to
look for common failures in comparison to related experimental data
that have become available recently \cite{NA49_new,NA49_Lam,Misch02} or
provide predictions for experimental studies in the near future
\cite{SPS20}, which are also of relevance for the new GSI-proposal
\cite{GSIprop}.

Our work is organized as follows:  In Section 2 we will describe the
main ingredients of the UrQMD and HSD transport approaches and point
out conceptual differences.  In Section 3 we study baryon stopping in
central $Au + Au$ collisions from 4 to 160 A$\cdot$GeV in comparison to
experimental data (whenever available). Section 4 is devoted to a
detailed comparison of both transport approaches on $\pi^\pm, K^+, K^-$
and $\Lambda+\Sigma^0$ rapidity distributions, yields  and different
particle ratios as a function of bombarding energy from 2 to 160
A$\cdot$GeV. Again the calculations will be confronted with
experimental data taken at the AGS and SPS. A direct comparison of
UrQMD and HSD on the $pp$ and $\pi^- p$ reaction level is given in
Section 5 to quantify the differences in the 'elementary' differential
cross sections. Section 6 concludes our study with a summary and
discussion of open problems.

\section{Transport models -- UrQMD and HSD}

In this work we employ two different transport models, i.e.\ the UrQMD
and HSD approaches, that have been used to described nucleus-nucleus
collisions from SIS to SPS energies for several years. Though different
in the numerical realisation, both models are based on the same
concepts: string, quark, diquark ($q, \bar{q}, qq, \bar{q}\bar{q}$) and
hadronic degrees of freedom. It is important to stress that both
approaches do not include any explicit phase transition to a
 quark-gluon plasma (QGP). The philosophy is that a common failure of
both models in comparison to experimental data should -- model
independently -- indicate the appearance of 'new physics'.

The UrQMD (Ultra-relativistic Quantum Molecular Dynamics) transport
approach is described in Refs.  \cite{UrQMD1,UrQMD2}. It includes all
baryonic resonances up to an invariant mass of 2 GeV as well as mesonic
resonances up to 1.9 GeV as tabulated in the PDG \cite{PDG}. For
hadronic continuum excitations a string model (let's denote it as
'Frankfurt' string model (FSM)) is used.  The hadron formation time
(which relates to the time between the formation  and fragmentation of
the string in the individual hadron-hadron center-of-mass frame) is in
the order of 1-2~fm/c depending on the momentum and energy of the
created hadrons (using the "yo-yo" formation concept for the time
definition) \cite{UrQMD1,UrQMD2}.  The UrQMD transport approach is
matched to reproduce nucleon-nucleon, meson-nucleon and meson-meson
cross section data in a wide kinematical regime \cite{UrQMD1,UrQMD2}.
At the high energies considered here the particles are essentially
produced in primary high energy collisions by string excitation and
decay, however, the secondary interactions among produced particles
(e.g.  pions, nucleons and excited baryonic and mesonic resonances) --
that also contribute to the particle dynamics -- are included as well.

Whereas UrQMD operates as default in the cascade mode, i.e. with hadron
potentials turned off, the HSD (Hadron-String Dynamics) transport
approach includes (by default) scalar and vector fields of the
particles which determine the mean-field propagation of the hadrons
between collisions (cf. Fig.~2 of Ref. \cite{HSD_exf}). The HSD
transport approach incorporates only the baryon octet and decuplet
states and N$^*$(1440), N$^*$(1535) as well as their antiparticles and
the $0^-$ and $1^-$ meson octets. Higher baryonic resonances are
discarded as explicit states (for propagation) in HSD; they are
supposed to "melt" in the nuclear medium even at normal nuclear density
(see e.g.\ \cite{meltB1,meltB2}). The argument here is that the
resonance structure (above the $\Delta$-peak) is not seen
experimentally even in photoabsorption on light nuclei \cite{gammaA}.
In contrast to the resonance concept -- adopted in UrQMD for all low
energy baryon-baryon and meson-baryon collisions -- HSD includes the
direct (non-resonant) meson production in order to describe the
corresponding cross sections (for the details see Ref.
\cite{CBRep98}).

In the HSD approach the high energy inelastic hadron-hadron collisions
are described by the LUND string model (realized by FRITIOF-7.02
\cite{LUND}), where two incoming hadrons emerge from the reaction as
two excited color singlet states, i.e.\ 'strings' (as in UrQMD).  The
formation time of all hadrons --- composed of light and strange quarks
--- in HSD is assumed to be $\tau_F \sim 0.8$ fm/c in the hadron rest
frame \cite{CBRep98,Ehehalt}, which is lower than the 'average' of the
exponentially distributed formation times of 1--2 fm/c used in UrQMD.
Note, that in  both models the formation time in the calculational
frame for heavy-ion collisions (laboratory or center-of-mass frame) is
dilated by the Lorentz $\gamma$-factor, i.e.\ $t_F=\gamma \cdot
\tau_F$.

Since at high energy heavy-ion collisions particle production
essentially proceeds via baryon-baryon and meson-baryon string excitations
and decays, it is worth to discuss the differences in the realizations of
the string models used in UrQMD and HSD.
In both string models the production probability $P$ of massive $s\bar{s}$
or $qq\bar{q}\bar{q}$ pairs is suppressed in comparison to light
flavor production ($u\bar{u}$, $d\bar{d}$) according to a generalized
Schwinger formula \cite{Schwinger}
\begin{eqnarray}
{P(s\bar{s}) \over P(u\bar{u})} = {P(s\bar{s}) \over P(d\bar{d})}=
\gamma_s = \exp\left(-\pi {m_s^2-m_q^2\over 2\kappa}, \right)
\label{schwinger}
\end{eqnarray}
with the string tension $\kappa\approx 1$~GeV/fm. Thus in the
string picture the production of strangeness and baryon-antibaryon
pairs is controlled by the masses of the constituent quarks and diquarks.
Inserting the constituent quark masses $m_u=0.3$~GeV and $m_s=0.5$ GeV
a value for the strangeness suppression factor $\gamma_s \approx 0.3$
is obtained. While the strangeness production in proton-proton
collisions at SPS energies is reasonably well reproduced in the LUND
string model with $\gamma_s=0.3$, the strangeness yield for $p+Be$
collisions at AGS energies (which is a good probe for the isospin
averaged elementary $p+p$ and $p+n$ reactions) is underestimated by
roughly 30\% \cite{Jgeiss}. Therefore the strangeness suppression
factor has been enhanced  to 0.4 at AGS energies for the elementary
nucleon-nucleon cross section in HSD. Thus, the
relative production probabilities for the different quark flavours
in the HSD model are
\begin{eqnarray}
{\rm HSD:} \ \ \ u : d : s: diquark = \left\{\begin{array}{l}
1 : 1 : 0.3 : 0.07 \; {\mathrm for }\; \sqrt{s}\ \geq\ 20\ {\mathrm GeV} \\
1 : 1 : 0.4 : 0.07 \; {\mathrm for }\; \sqrt{s}\ \leq\ 5\ {\mathrm GeV}
\end{array} \right.
\label{supfHSD}\end{eqnarray}
with a linear transition  of the strangeness suppression factor as a
function of $\sqrt{s}$ in between.
The relative production probabilities for the different quark flavors
in UrQMD are fitted to
\begin{eqnarray}
{\rm UrQMD:} \ \ \ u : d : s: diquark =
1 : 1 : 0.35 : 0.1.
\label{supfUr}\end{eqnarray}
Additionally fragmentation functions $f(x,m_t)$ must
be specified, which are the probability distributions for hadrons
with transverse mass $m_t$ to acquire the energy-momentum fraction
$x$ of the fragmenting string.
One of the most common fragmentation functions is used in the LUND model
\cite{LUND} (which is adopted in the HSD approach \cite{Jgeiss})
\begin{eqnarray}
f(x,m_t)\approx {1 \over x} (1-x)^a \exp\left(-bm_t^2/x  \right),
\label{fragf}
\end{eqnarray}
with $a=0.23$ and $b=0.34$~GeV$^{-2}$.
In UrQMD different fragmentation functions are used for leading nucleons
and newly produced particles, respectively (cf. Ref. \cite{UrQMD1}
Fig.\ 3.16):
\begin{eqnarray}
&& f(x)_{\mathrm nucl} = \exp\left(-{(x-B)^2\over 2 A}\right),
        \ {\rm for \ leading \ nucleons} \nonumber\\
&& f(x)_{\mathrm prod} = (1-x)^2,\ {\rm for \ produced \ particles}
\label{fragfUr}
\end{eqnarray}
with $A=0.275$ and $B=0.42$. The fragmentation function $f(x)_{\mathrm
prod}$ --- used for newly produced particles --- is the well-known
Field-Feynman fragmentation function \cite{frfanUr}. At the string
break-up the $q\bar q$-pairs have zero transverse momenta in the string
reference frame, but the transverse momentum distributions of the
single quark ($\vec p_t$) and the corresponding antiquark ($-\vec p_t$)
are taken to be gaussian
\begin{eqnarray}
f(p_t) = {1\over \sqrt{\pi\sigma^2}} \exp\left(-{p_t^2\over \sigma^2}\right)
\label{fragfUrpt}
\end{eqnarray}
with $\sigma=1.6$~GeV/c.

Despite the differences in the fragmentation functions, both
string models describe quite well the data available for particle
multiplicities and total spectra from $pp$ collisions at high energies
(see Ref. \cite{UrQMD1} for UrQMD and Ref. \cite{Jgeiss} for HSD).
Also the inelastic pion-proton cross section is in good agreement with
the experimental data in both models whereas differential spectra can
differ substantially (cf. Section V).  The LUND string model (in HSD)
has also been tested for low energy $pp$ collisions as well as for $\pi
N$ interactions (cf. Chapter 2 in Ref. \cite{Jgeiss}). It has been
shown that the (LUND) string model underestimates pion and
kaon/antikaon yields closer to their production threshold. In HSD the
threshold for string formation and decay thus is taken as $\sqrt{s}$ =
2.6~GeV for baryon-baryon collisions and at $\sqrt{s}$ = 2.1~GeV for
meson-baryon collisions. For lower invariant energies $\sqrt{s}$
resonant and direct meson production mechanisms (e.g.\ $\pi N\to N
\pi\pi$) dominate, which are implemented in addition in HSD to ensure
smooth excitation functions of the meson multiplicities from threshold
to a few hundred GeV/c.

\section{Baryon stopping}

\begin{figure}[t]
\phantom{a}\vspace*{-1.5cm}
\centerline{\psfig{figure=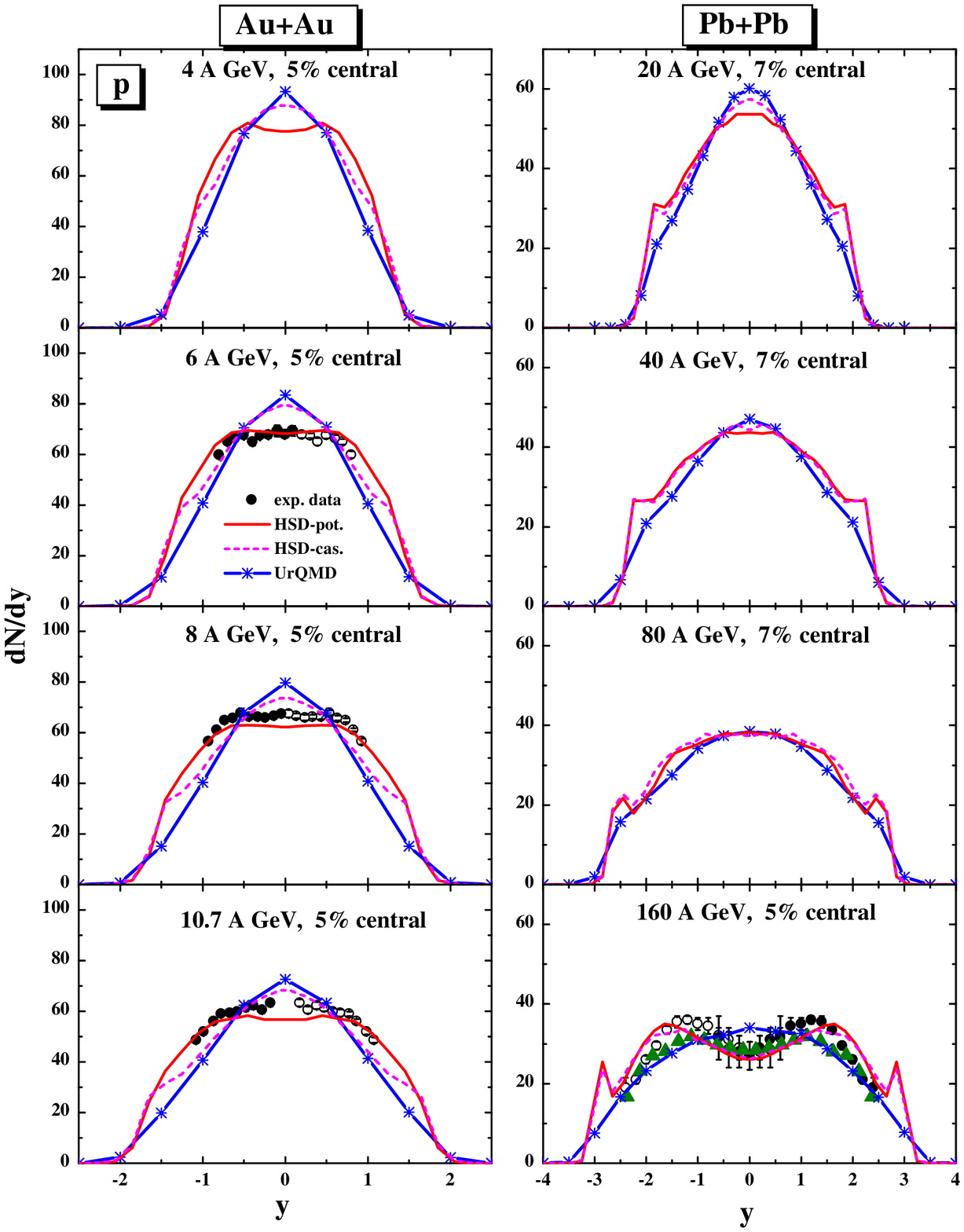,width=14cm}}
\vspace*{5mm}
\caption{The rapidity distributions for protons from 5\% (4, 6, 8, 10.7
and 160 A$\cdot$GeV) and 7\% central (20, 40, 80 A$\cdot$GeV) $Au+Au$
(AGS) and $Pb+Pb$ (SPS) collisions at 4--160 A$\cdot$GeV.
The experimental data at 4, 6, 8, 10.7 A$\cdot$GeV have been taken from
Ref.~\protect\cite{E917p02} (circles), at 160 A$\cdot$GeV from
\protect\cite{NA49pNew} (triangles) and from \protect\cite{NA49pold}
(circles).  The full symbols correspond to the measured data, whereas
the open symbols are the data reflected at midrapidity.  The solid
lines with stars show the results from the UrQMD calculations while the
solid and dashed lines stem from the HSD approach with and without
potentials, respectively.}
\label{Fig1}
\end{figure}

Though various predictions have been made in both transport models
since a couple of years, it is of importance to compare with actual
data. We employ the experimental cuts in centrality to get as realistic
as possible in the comparison of baryon stopping achieved in both theoretical
approaches and in the different experiments. A related comparison is
presented in Fig.~\ref{Fig1} for protons from 5\% (4, 6, 8, 10.7 and
160 A$\cdot$GeV) and 7\% central (20, 40, 80 A$\cdot$GeV) $Au+Au$ (AGS)
and $Pb+Pb$ (SPS) collisions at 4--160 A$\cdot$GeV \footnote{Note, that
for all UrQMD and HSD calculations presented in this work the
centrality of the reaction has been determined by a comparison of the
transport calculations to the energy distribution in the
Veto-calorimeter of the NA49 collaboration for SPS energies and to the
New Multiplicity Array (NMA) and ZCAL-calorimeters at AGS energies}.
The experimental data at 4, 6, 8, 10.7 A$\cdot$GeV have been taken from
Ref.~\cite{E917p02} (circles), at 160 A$\cdot$GeV from \cite{NA49pNew}
(triangles) and from \cite{NA49pold} (circles).  The full symbols (here
and for all further figures) correspond to the measured data whereas
the open symbols are the data reflected at midrapidity.  The solid
lines with stars show the results from the UrQMD calculations while the
solid and dashed lines stem from the HSD approach with and without
potentials, respectively.

We note, that in the UrQMD calculations 'spectator' protons have been
cut off whereas they are still present in the HSD calculations; this
leads to the maxima in the proton rapidity distributions at target and
projectile rapidity in the HSD calculations (cf. Fig.~1).
Nevertheless, the HSD cascade calculations are found to agree with
UrQMD cascade calculations from 4--20 A$\cdot$GeV within 5\%, whereas
UrQMD shows somewhat more proton stopping than HSD at higher bombarding
energies. The mean-field propagation effects in the HSD approach are
most pronounced at low bombarding energies leading to a reduction of
baryon stopping from 4 -- 10.7 A$\cdot$GeV and a flatter rapidity
distribution $dN/dy$ around midrapidity slightly closer to the
experimental data. This effect can easily be attributed to the energy
stored in the repulsive mean field at high baryon density and moderate
bombarding energy.  Above about 40 A$\cdot$GeV such potential effects
are no longer statistically significant in the calculations since the
repulsive mean field decreases strongly with momentum (cf. Fig.~2 of
Ref. \cite{HSD_exf}) such that at 40 A$\cdot$GeV no repulsion is seen
by the nucleons in the initial high density phase. Only when the
system partly thermalizes and the nucleon momenta relative to the
fireball reference frame become smaller the baryons 'feel' again a
repulsive mean field, however, now at rather low baryon density. We
recall that at density $\rho_0$ the potential is even attractive for
momenta $\leq$ 600 MeV/c.

In general the HSD results indicate slightly less baryon stopping than
the UrQMD calculations.  At 160 A$\cdot$GeV the experimental data favor
a minimum of the distribution at midrapidity, which is reproduced by
the HSD calculations. However, the UrQMD calculations only deviate by
$\sim$ 5\%. We note, that for semi-central and peripheral $Pb + Pb$
collisions at 160 A$\cdot$GeV the UrQMD calculations are in good
agreement with the data from Ref. \cite{NA49pNew} (cf. Fig.~1 in
Ref. \cite{WBH_stop02}).

Thus, the overall description of the rapidity spectra from  both models
(with/without potentials) in this wide energy regime is quite
remarkable in view of the different 'hadronic' degrees of freedom and
string 'parameters' involved.

\section{Pion and strangeness production}

\begin{figure}[t]
\centerline{\psfig{figure=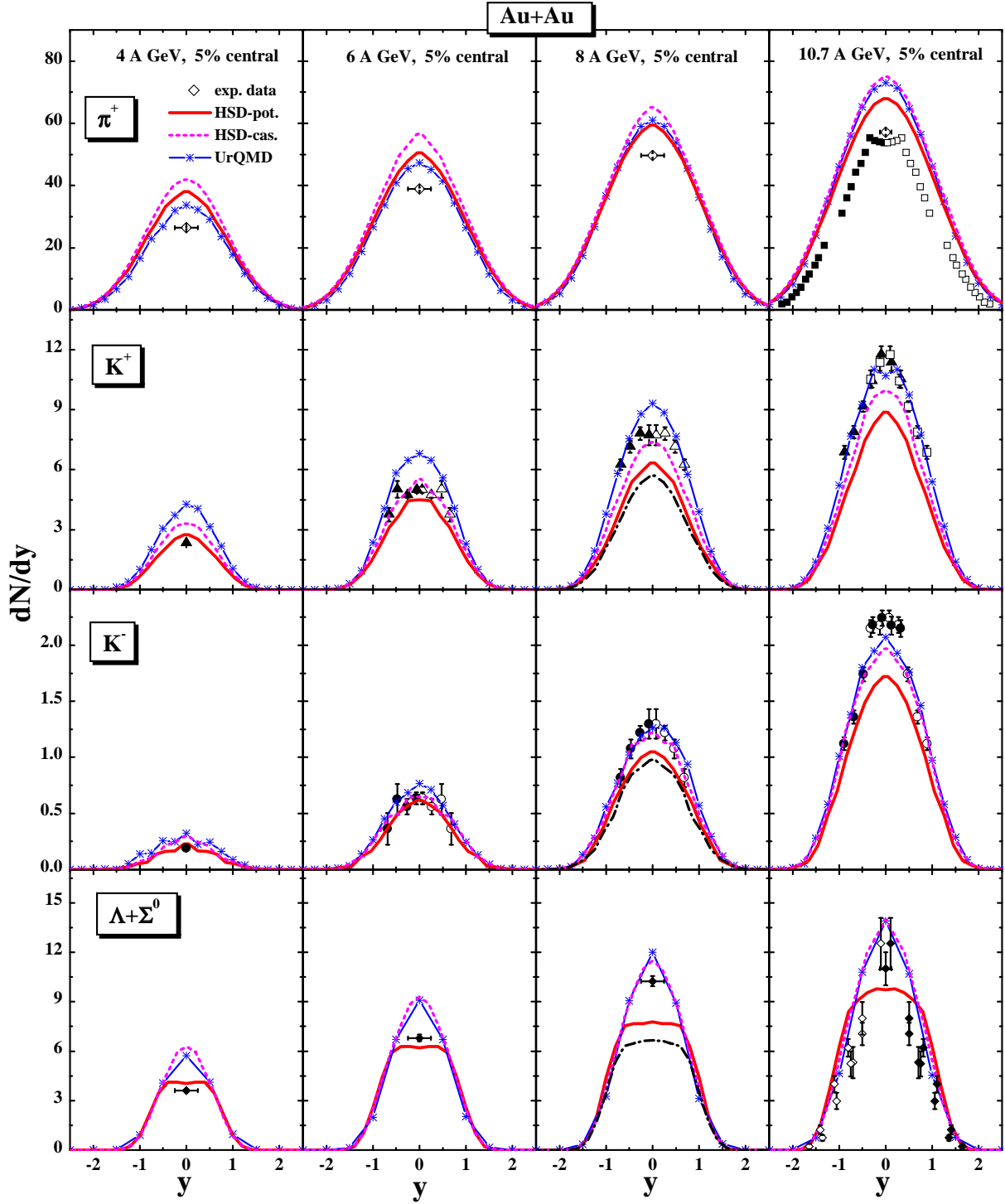,width=16cm}}
\vspace*{5mm}
\caption{The rapidity distributions for $\pi^+, K^+, K^-$ and
$\Lambda+\Sigma^0$'s from 5\% central $Au+Au$ collisions at 4--10.7
A$\cdot$GeV in comparison to the experimental data from Refs.
\protect\cite{E866E917,E917K,E866pi11,E877pi11,E891Lam,E896Lam}.
The thin solid lines with stars show the results from the UrQMD
calculations while the thick solid and dashed lines stem from the HSD
approach with and without potentials, respectively.}
\label{Fig2}
\end{figure}

\begin{figure}[t]
\centerline{\psfig{figure=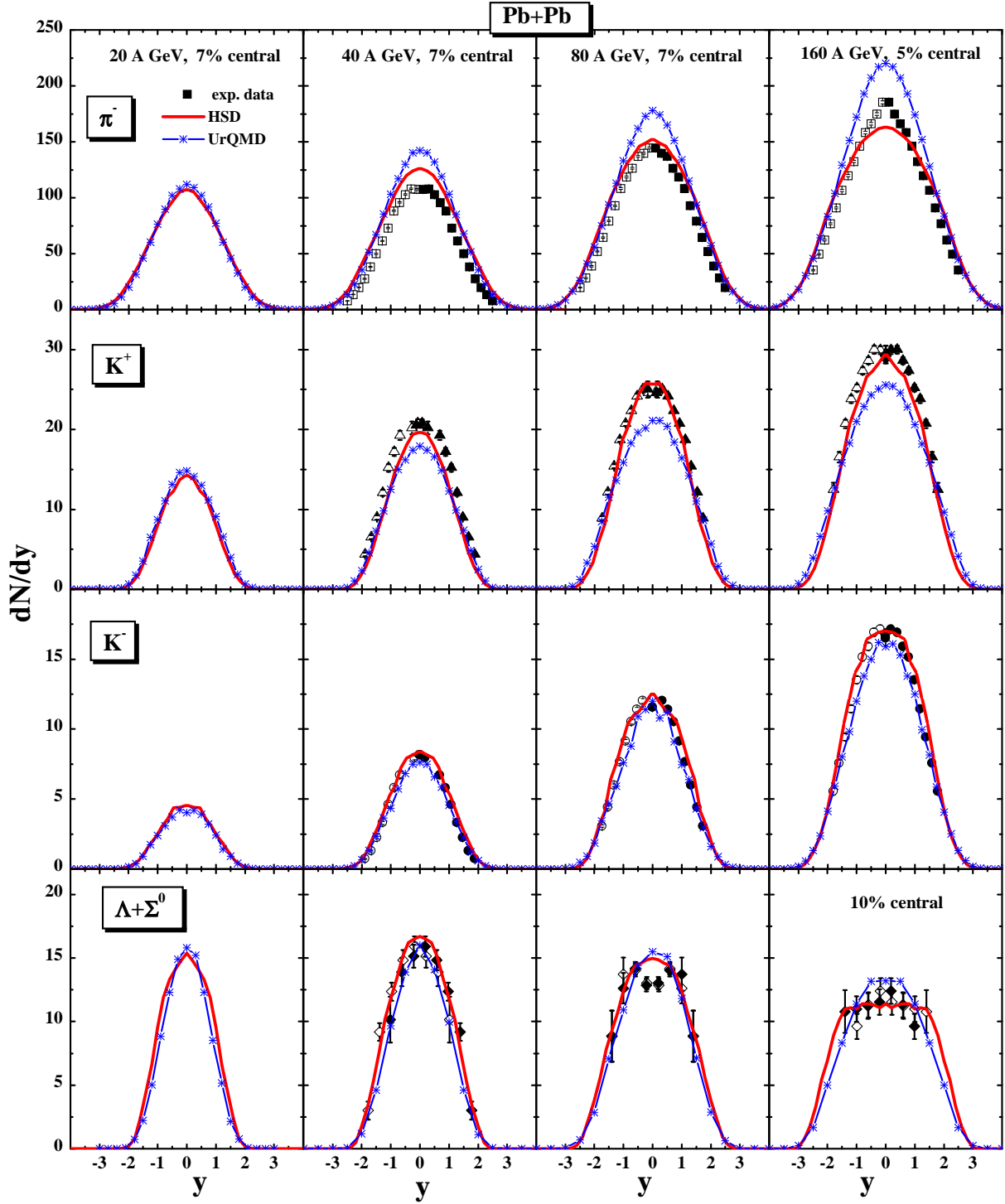,width=16cm}}
\vspace*{5mm}
\caption{The rapidity distributions for $\pi^-, K^+, K^-$ and $\Lambda
+\Sigma^0$'s from 5\% (160 A$\cdot$GeV), 7\% (20, 40 and 80
A$\cdot$GeV) or 10\% central ($\Lambda+\Sigma^0$ at 160 A$\cdot$GeV)
$Pb+Pb$ collisions at 20--160 A$\cdot$GeV in comparison to the
experimental data from Refs. \protect\cite{NA49_new,NA49_Lam,Misch02}. The
solid lines with stars show the results from the UrQMD calculations
while the thick solid and dashed lines stem from the HSD approach with
and without potentials, respectively.}
\label{Fig3}
\end{figure}

We continue with $\pi^{\pm}, K^+, K^-$ and $\Lambda+\Sigma^0$ rapidity
spectra in 5\%, 7\%, or 10\% central collisions of $Au+Au$ or $Pb+Pb$,
respectively, from 4--160 A$\cdot$GeV. We compare to the AGS
data from Refs.
\cite{E866E917,E917K,E866pi11,E877pi11,E891Lam,E896Lam} in Fig. \ref{Fig2}
and SPS data  from Refs. \cite{NA49_new,NA49_Lam,Misch02} in Fig. \ref{Fig3}.
Here the thick solid lines denote the HSD results including the
potentials, the dashed lines represent HSD calculations in the cascade
mode, which should be directly compared to the UrQMD results (thin solid
lines with stars). Both Figs. -- taken together -- provide an overview
on the energy dependence of the different rapidity distributions and
the virtues/failures of the transport models.

At 4 A$\cdot$GeV all transport versions overestimate the $\pi^+$
spectra. This deviation is most pronounced in the HSD 'cascade'
version. On the other hand, UrQMD is higher than HSD in the strangeness
channels $K^\pm$ and $\Lambda+\Sigma^0$, whereas HSD (with potentials)
is quite compatible with the midrapidity data. At 6 A$\cdot$GeV the HSD
(with potential) calculations are in line with the experimental $K^\pm$
and $\Lambda+\Sigma^0$ data whereas the UrQMD results are still too
high. All calculations again overestimate the $\pi^+$ rapidity spectra.
At 8 A$\cdot$GeV the UrQMD and HSD results continue to overestimate the
$\pi^+$ yield, while now the HSD calculations (with potential) fall low
in the strangeness channels contrary to the HSD cascade and UrQMD
results. At 10.7 A$\cdot$GeV the picture continues:  both models give
too many $\pi^+$, the strangeness yield is approximately reproduced in
the cascade versions and underestimated in the HSD simulation with
potentials.

At 20 A$\cdot$GeV, most relevant for the future GSI facility
\cite{GSIprop}, both transport models (with/without potentials) give
the same rapidity distributions for all hadrons considered here. Note,
that the cascade HSD results are not shown explicitly in Fig.~\ref{Fig3}
since the cascade and potential results are practically
coincident. As discussed above this is due to the momentum-dependence
of the scalar and vector potentials in HSD; both potentials vanish for relative
momenta of a couple of GeV/c. Deviations between HSD and UrQMD come up again at 40
A$\cdot$GeV, where UrQMD now gives more $\pi^-$ and lacks a few $K^+$,
whereas the $K^-$ and $\Lambda+\Sigma^0$ spectra are reasonably
reproduced. At this energy HSD (with/without potentials) only
overestimates the $\pi^-$ distribution slightly.  At 80 A$\cdot$GeV the
UrQMD spectra are also fine for $K^-$ and $\Lambda$, but low for $K^+$
and too high for $\pi^-$. At 160 A$\cdot$GeV the HSD calculations
indicate slightly too few $\pi^-$ at midrapidity, however, do well
for $K^\pm$ and $\Lambda+\Sigma^0$ as already demonstrated earlier in Refs.
\cite{CBRep98,Jgeiss}.  The UrQMD results here show too many $\pi^+$
and a slightly higher $\Lambda+\Sigma^0$ yield, whereas the $K^+$
distribution is underestimated. We note, that the UrQMD results
presented in Fig.~\ref{Fig3} are taken from Ref.~\cite{WBS_K02}.

\begin{figure}[t]
\centerline{\psfig{figure=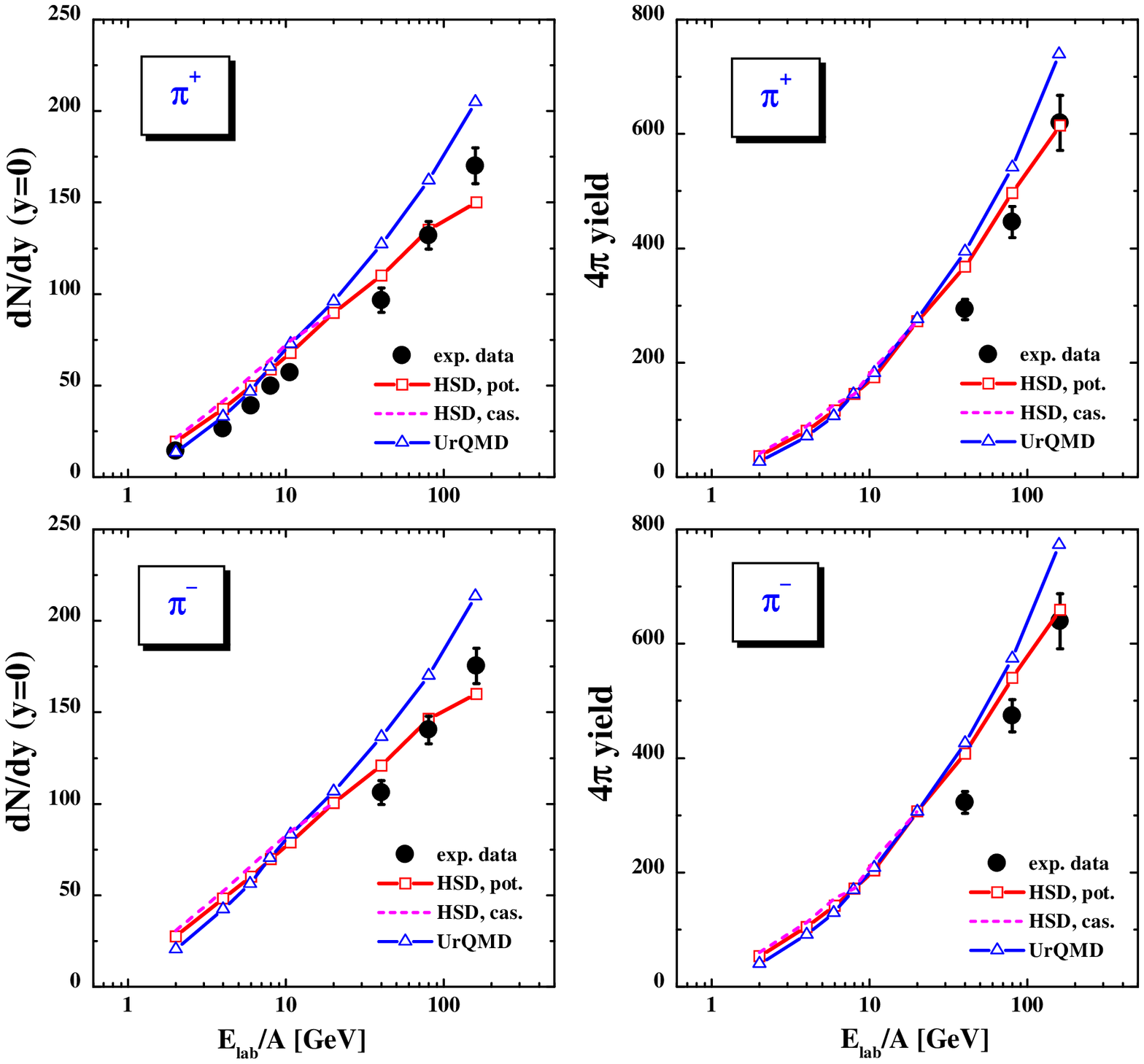,width=16cm}}
\vspace*{5mm}
\caption{The excitation function of $\pi^+$ (upper plots) and $\pi^-$
(lower plots) yields from 5\% (AGS energies and 160 A$\cdot$GeV)
or 7\% central (20, 40 and 80 A$\cdot$GeV) $Au+Au$ (AGS) or
$Pb+Pb$ (SPS) collisions  in comparison to the experimental data from Refs.
\protect\cite{E866E917,NA49_new} for midrapidity (left column) and
rapidity integrated yields (right column).
The solid lines with open triangles show the results from the UrQMD
calculations while the solid lines with open squares and dashed lines
stem from the HSD approach with and without potentials, respectively. }
\label{Fig4}
\end{figure}

As a more general overview on the $\pi^\pm$ abundancies in central
$Au+Au$ and $Pb+Pb$ collisions we show in Fig.~\ref{Fig4}
the $\pi^+$ (upper plots) and $\pi^-$ (lower plots) multiplicities
at midrapidity (left column) and integrated over rapidity (right
column) as a function of the bombarding energy in
comparison to the available data from Refs.  \cite{NA49_new,E866E917}.
Here the solid lines with open triangles show the results from the UrQMD
calculations while the solid lines with open squares and dashed lines
stem from the HSD approach with and without potentials, respectively.
At lower AGS energies the UrQMD model gives slightly less pions then
HSD (with/without potential), but  both models overpredict the
midrapidity data (except UrQMD at 2 A$\cdot$GeV, which is in line with
data point at midrapidity). About 20 A$\cdot$GeV is the "crossing
point" for both transport calculations and at SPS energies the tendency
turns around: UrQMD gives more pions than HSD, so that HSD is now in a
better agreement with the experimental data.

The overestimation of the pion yield at low and high energies by
the transport models is presently not well understood. In Ref.
\cite{larionov} Larionov et al. have speculated that higher baryon
resonances -- more massive than the $\Delta$-resonance -- might be
quenched at densities  above $\sim$ 1.5 $\rho_0$ both in $NN$ as
well as $\pi N$ channels. The net effect at SIS energies (and
slightly higher bombarding energies) is a reduction of the pion
yield in collisions of heavy systems such as $Au+Au$ whereas light
systems (e.g. $C+C$) are not effected very much \cite{larionov}.
Though this might be a possible explanation at SIS and lower AGS
energies, other effects such as strong pion self energies could
also lead to a reduction of the pion abundance in the transport
models. Such many-body effects have not been incorporated in the
calculations presented here. At SPS energies and especially at 160
A$\cdot$GeV another production channel for pions becomes sizeable,
i.e. the annihilation of baryon-antibaryon ($B\bar{B}$) pairs that
leads on average to 5 pions (or more). However, by employing
detailed balance on the many-body level \cite{Cass02_antip} the
$B\bar{B}$ annihilation rate is found to be almost compensated by
many-meson fusion channels that recreate $B\bar{B}$ pairs. This
many-body channel is not incorporated in UrQMD
and thus might partly be responsible for the overestimation of the
pion yield. A rough estimate, however, shows that this missing
channel cannot be the only reason: as calculated in Ref.
\cite{Cass02_antip} (and approved by recent NA49 measurements
\cite{Misch02}) the $\bar{p}/<\pi>$ ratio ($<\pi> = 0.5 (\pi^+ +
\pi^-)$) is approximately independent on the centrality of the $Au+Au$
collision and $\sim$ 1\%. Assuming the number of $\bar{n}$ to be equal
to the number of antiprotons as well as the number of antihyperons
\cite{Misch02} and adopting isospin symmetry for the pions, we get
$\bar{N}/\pi \approx $ 1 \%. If all $\bar{N}$'s produced initially are
annihilated, which is an upper estimate for the pion production from
this channel, then we could achieve a maximum increase of the pion
number by 5 \%.  Thus a $\sim$20 \% overestimation of the pion yield by
UrQMD at 160 A$\cdot$GeV can only partly be attributed to the missing
multi-meson fusion channels.

\begin{figure}[t]
\centerline{\psfig{figure=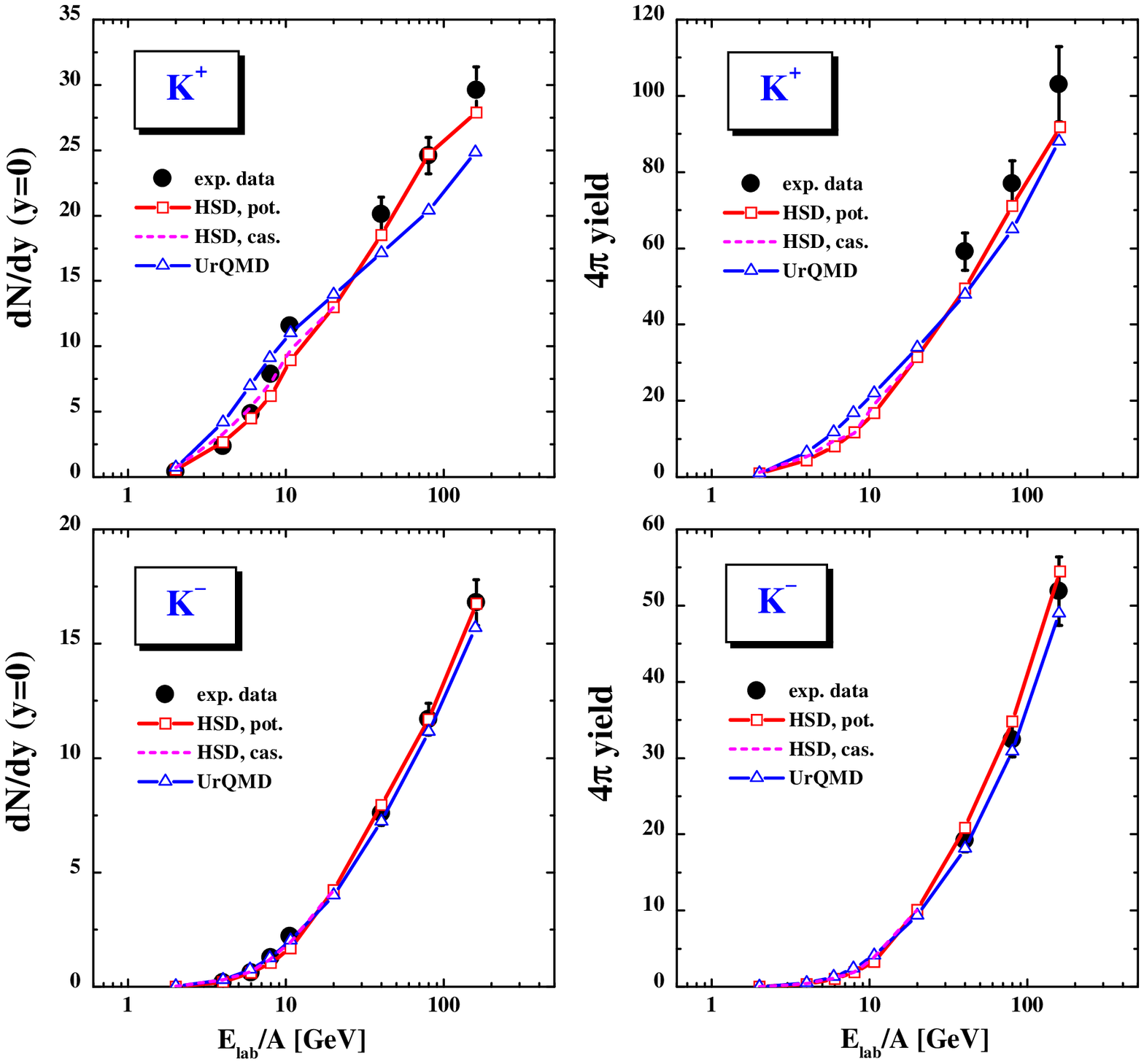,width=16cm}}
\vspace*{5mm}
\caption{The excitation function of $K^+$ (upper plots) and $K^-$
(lower plots) yields from 5\% (AGS energies and 160 A$\cdot$GeV) or 7\%
central (20, 40 and 80 A$\cdot$GeV) $Au+Au$ (AGS) or $Pb+Pb$ (SPS)
collisions  in comparison to the experimental data from Refs.
\protect\cite{E866E917,NA49_new} for midrapidity (left column) and
rapidity integrated yields (right column).
The solid lines with open triangles show the results from the UrQMD
calculations while the solid lines with open squares and dashed lines
stem from the HSD approach with and without potentials, respectively. }
\label{Fig5}
\end{figure}

The $K^+$ (upper plots) and $K^-$ (lower plots) multiplicities at
midrapidity (left column) and integrated over rapidity (right column)
are shown in Fig.~\ref{Fig5} as a function of the bombarding energy in
comparison to the available data from Refs. \cite{NA49_new,E866E917}.
Here the midrapidity and total yields summarize the findings from Figs.
\ref{Fig2}-\ref{Fig3}: The $K^-$ abundancies are well described by both
transport models. The $K^+$ yield at midrapidity (left column) is slightly
overestimated by UrQMD at AGS energies and underestimated at SPS
energies, whereas HSD is in a reasonable agreement with the data except
of the upper AGS energies (with potentials included). This tendency
stays the same for the $4\pi$ kaon yield, however, at SPS energies
the difference between both models for the $4\pi$ yields is smaller than at
midrapidity since UrQMD provides a slightly broader kaon rapidity
distribution than HSD (cf. Fig.\ \ref{Fig3}).  Thus, an underestimation
of strangeness production is not the prevailing issue as demonstrated
in Fig.\ \ref{Fig5} in comparison to the recent experimental data from
NA49 \cite{NA49_new}. Both transport models can roughly describe -
within their systematic range of uncertainties - the $K^\pm$ spectra
and abundancies.

\begin{figure}[t]
\centerline{\psfig{figure=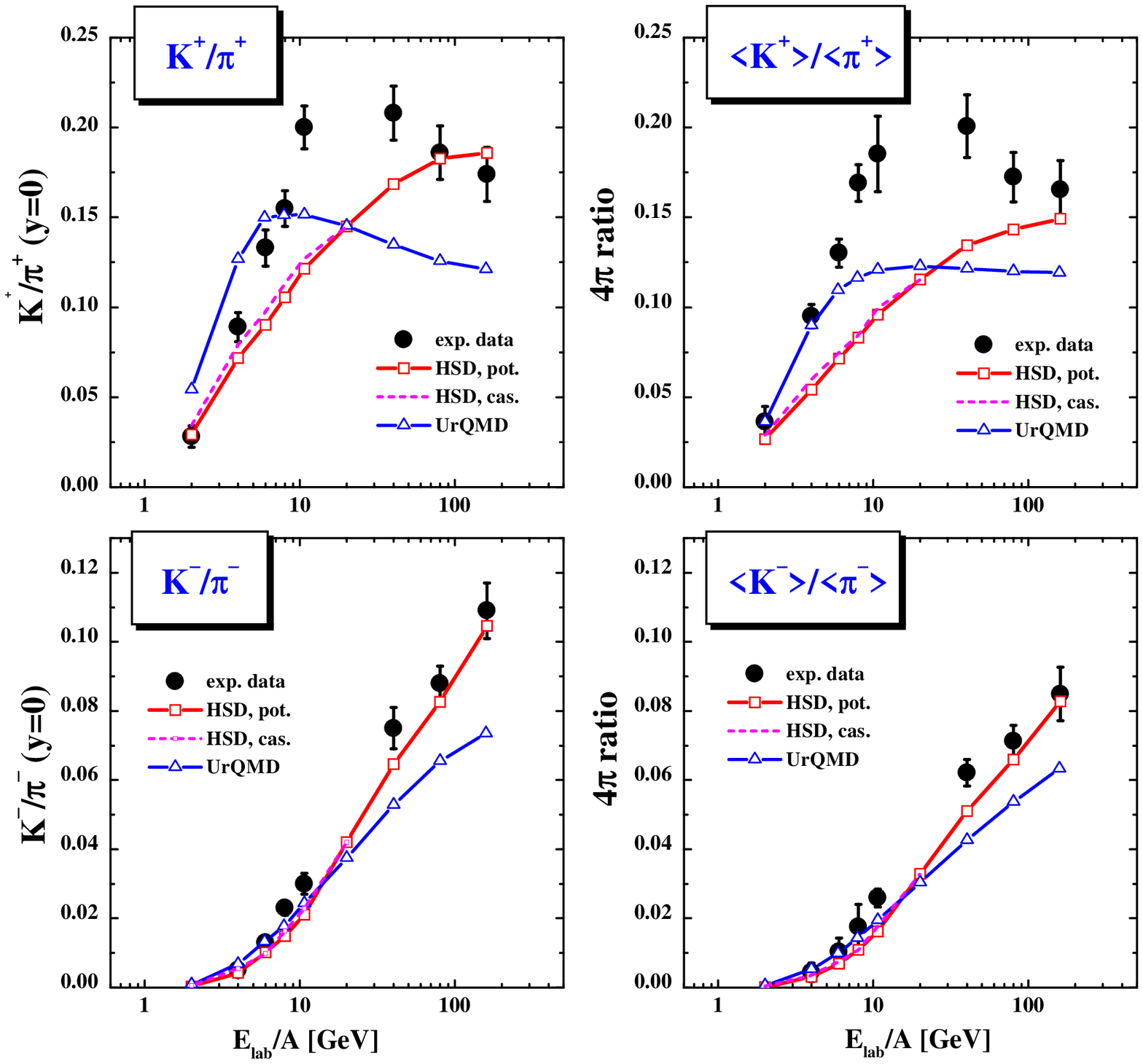,width=16cm}}
\vspace*{5mm}
\caption{The excitation function of the $K^+/\pi^+$ (upper plots) and
$K^-/\pi^-$ (lower plots) ratios from 5\% (AGS energies and 160
A$\cdot$GeV) or 7\% central (20, 40 and 80 A$\cdot$GeV) $Au+Au$ (AGS)
or $Pb+Pb$ (SPS) collisions  in comparison to the experimental data from
Refs. \protect\cite{E866E917,NA49_new} for midrapidity (left
column) and rapidity integrated yields (right column).
The solid lines with open triangles show the results from the UrQMD
calculations while the solid lines with open squares and dashed lines
stem from the HSD approach with and without potentials, respectively. }
\label{Fig6}
\end{figure}

In Fig.\ \ref{Fig6} we show the $K^+/\pi^+$ and $K^-/\pi^-$ ratios at
midrapidity (left column) and integrated over all angles (right column)
as a function of the bombarding energy for central collisions of
$Au+Au$ (AGS) or $Pb+Pb$ (SPS) in comparison to the available data from
Refs. \cite{NA49_new,E866E917}.  Whereas the excitation function of the
$K^-/\pi^-$ ratio is roughly reproduced by both transport models, the
maximum in the $K^+/\pi^+$ ratio seen experimentally both at
midrapidity (upper left part) and in $4\pi$ (upper right part) is not
described by HSD as well as UrQMD. For the $K^+/\pi^+$ ratio both
models give quite different results. HSD gives a monotonous increase of
this ratio with bombarding energy (as pointed out in Refs.
\cite{HSD_exf,Jgeiss} \footnote{We note that the HSD results presented
in this work are produced with much higher statistics than in Refs.
\cite{HSD_exf,Jgeiss} due to the increasing computer power available.
Also the centrality selection is done now in line with the actual
experimental set-up.}), whereas within UrQMD the ratio shows a maximum
around 10 A$\cdot$GeV and then drops slightly for the midrapidity ratio
or stays roughly constant for the $4\pi$ ratios. In view of Figs.\
\ref{Fig2}-\ref{Fig5} this failure is not primarily due to a mismatch
of strangeness production, but more due to an insufficient description
of the pion abundancies.

\begin{figure}[t]
\centerline{\psfig{figure=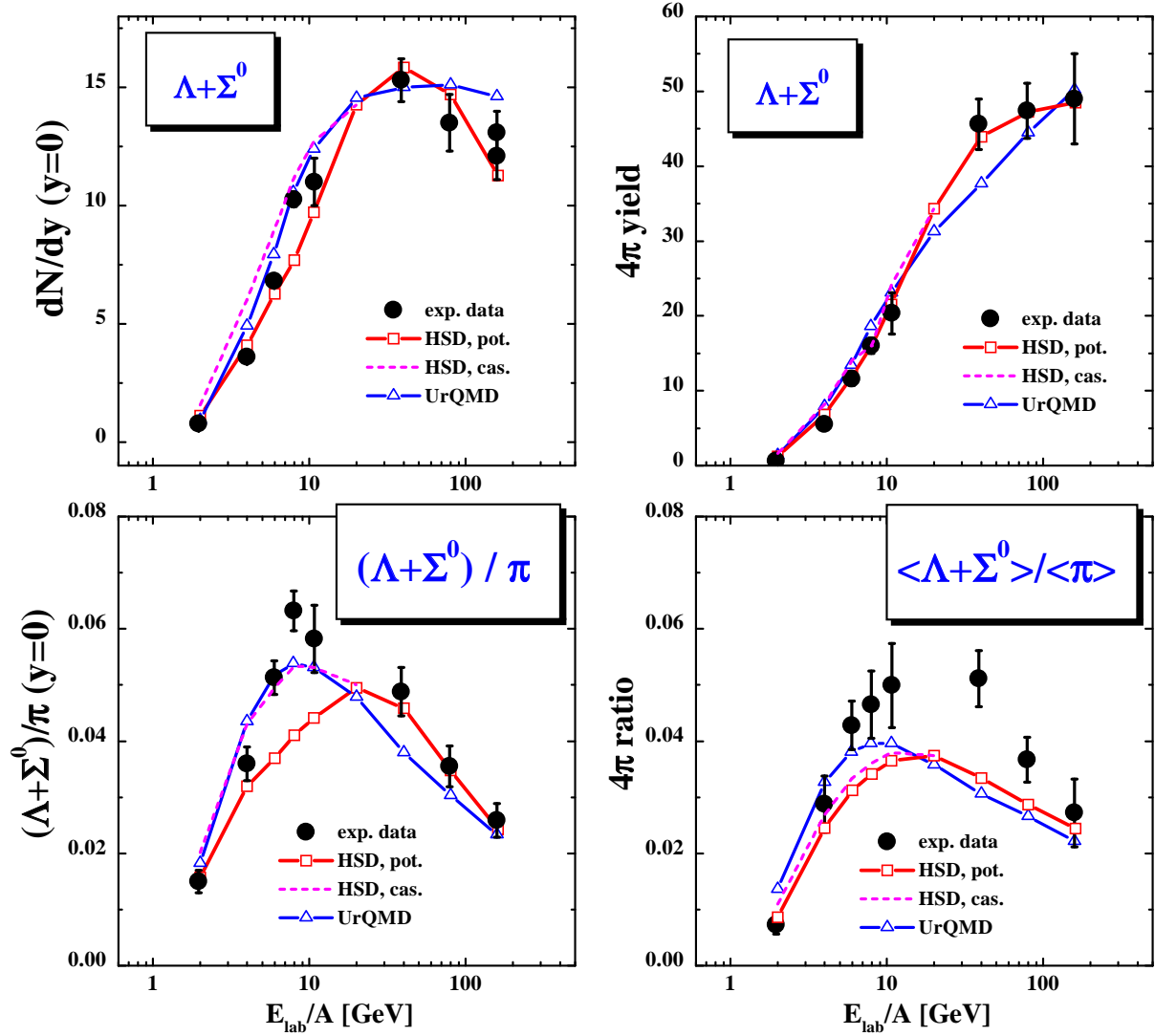,width=16cm}}
\vspace*{5mm}
\caption{The excitation function of the $\Lambda+\Sigma^0$ yields and
$(\Lambda+\Sigma^0)/\pi$ ratio from 5\% (AGS energies), 7\% (20, 40 and 80
A$\cdot$GeV) or 10\% central (160 A$\cdot$GeV) $Au+Au$ (AGS) or $Pb+Pb$
(SPS) collisions in comparison to the experimental data from Refs.
\protect\cite{NA49_Lam,Misch02,E891Lam,Antiori} for midrapidity (left column) and rapidity
integrated yields (right column).
The solid lines with open triangles show the results from the UrQMD
calculations while the solid lines with open squares and dashed lines
stem from the HSD approach with and without potentials, respectively. }
\label{Fig7}
\end{figure}

Fig.\ \ref{Fig7} shows the excitation functions of
$\Lambda+\Sigma^0$-hyperons at midrapidity (upper left part) and
integrated over all angles (upper right part) as a function of the
bombarding energy for central collisions of $Au+Au$ (AGS) or $Pb+Pb$
(SPS) in comparison to the available data from Refs.
\cite{NA49_Lam,Misch02,E891Lam,E896Lam,Antiori}.  Here UrQMD (solid
lines with open triangles) slightly underestimates the $4\pi$ yields at
40 and 80 A$\cdot$GeV whereas HSD (solid lines with open squares) seems
to give a better description.  Nevertheless, all models compare rather
well with data.

The $(\Lambda+\Sigma^0)/\pi$ ratios\footnote{The pion multiplicity is
calculated as $\pi=3/2(\pi^+ +\pi^-)$ in line with
Ref.~\cite{NA49_Lam,Misch02}} at midrapidity (lower left part) and
integrated over $4\pi$ (lower right part) are underestimated slightly
which should again be attributed to the pion excess in the transport
models (see above).  Nevertheless, the maxima in the ratios ($4\pi$ and
midrapidity) observed experimentally is qualitatively reproduced by
both models indicating that with increasing bombarding energy
$s$-quarks are more frequently produced within mesons ($\bar{K},
\bar{K}^*$) rather than in associate production with baryons. A similar
trend is also found in statistical model fits \cite{StatMod}.

\begin{figure}[t]
\centerline{\psfig{figure=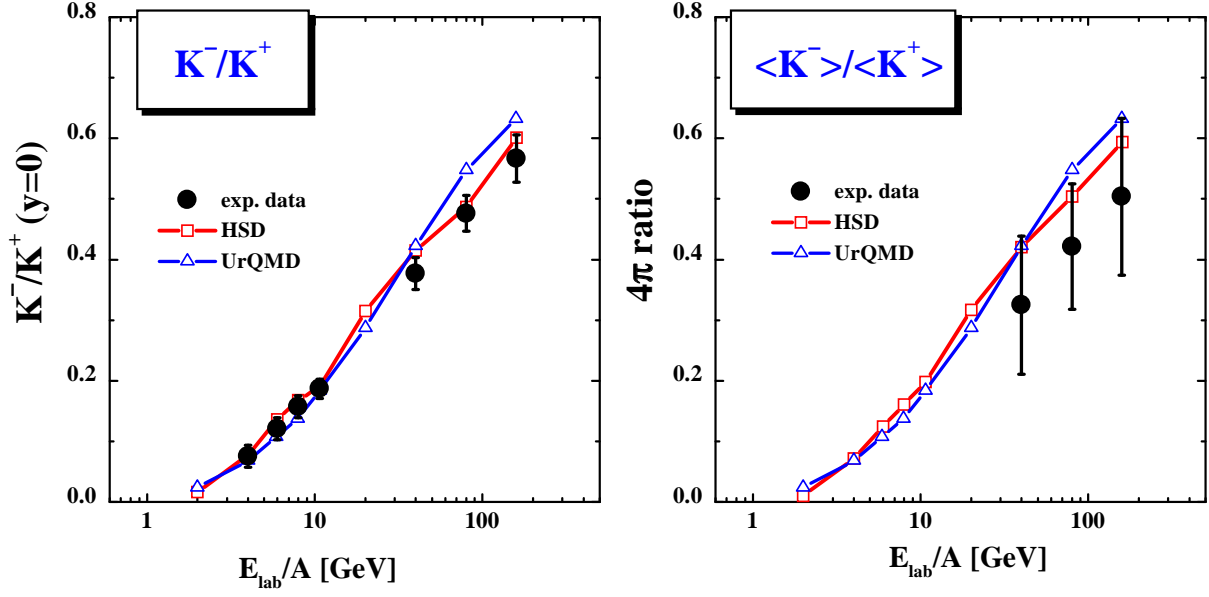,width=16cm}}
\vspace*{5mm}
\caption{The excitation function of the $K^-/K^+$ ratio from
5\% (AGS energies and 160 A$\cdot$GeV) or 7\% central
(20, 40 and 80 A$\cdot$GeV) $Au+Au$ (AGS) or $Pb+Pb$ (SPS) collisions
in comparison to the experimental data from Refs.
\protect\cite{E866E917,NA49_new} for midrapidity (left
column) and rapidity integrated yields (right column).
The solid lines with open triangles show the results from the UrQMD
calculations while the solid lines with open squares
stem from the HSD approach with potentials. }
\label{Fig8}
\end{figure}

The excitation function for the $K^-/K^+$ ratio in central collisions
of $Au+Au$ (AGS) or $Pb+Pb$ (SPS) is shown in Fig.\ \ref{Fig8} for
midrapidity ratios (l.h.s.) and in $4\pi$ (r.h.s.).  Experimental data
\cite{NA49_new,E866E917} here are only available for the midrapidity
ratios.  Again we find that both transport models give similar results
for this ratio which are comparable to the data.  Statistical models
also fit this ratio quite well. Whether this finding implies that
'chemical equilibration' is approximately achieved in central
collisions of heavy nuclei is still an open question
\cite{Zschiesche:2002zr}.

One note of caution again:
the pions in both transport models are treated as 'free' particles,
i.e. with their vacuum mass. On the other hand, lattice QCD
calculations as well  as effective Lagrangian approaches like the
Nambu-Jona-Lasinio (NJL) model show that the pion mass should increase
with temperature and density. So the overestimation of the pion yields
in HSD and UrQMD might be a signature for a dynamically larger pion
mass. Moreover, in-medium changes of the strange hadron properties, as
known from experimental studies at SIS energies, should also show up at
AGS and SPS energies. Thus, including all medium effects simultaneously
in a consistent way might provide a more conclusive interpretation of
the ratios in Figs.\ \ref{Fig6}-\ref{Fig8}.  However, such calculations
require a precise knowledge about the momentum and density dependence
of the hadron self-energies which is not available so far.  Note, that
up to now in-medium modifications of the $K^+, K^-$ properties have
been studied with HSD employing a dropping of $K^-$ and increase of
$K^+$ masses in the medium \cite{Kmdrop}.  As summarized in
\cite{CBRep98} such a scenario leads to an enhancement of $K^-$ and a
lowering of $K^+$ yields at SIS energies (which is close to threshold
for $K^+, K^-$ production). It modifies only slightly the strangeness
abundancies at SPS energies. However, chiral
symmetry restoration also requires a simultaneous modification of the pion
properties.

We close this Section with some speculations about the failure of
the transport models for the $K^+/\pi^+$ ratio at the top AGS
energies (and slightly above). To this aim let's assume that the
pion yield is decreased by some mechanism to the actual yield
observed experimentally. Since kaons and antikaons are also
produced in secondary non-strange meson-baryon collisions, this
will imply also a reduction of the kaon number in the transport
calculations. For central $Au+Au$ reactions these secondary
channels give roughly the same amount of kaons and antikaons as
the primary $NN$ channels. Thus a 20\% decrease in the pion number
will lead to a maximum change in the kaon number by 10\% and the
$K^+/\pi^+$ ratio might increase by about 10\%, too. This relative
increase will improve the situation in comparison to experiment,
however, not solve the problem. Another possibility is the
enhanced production of $s\bar{s}$ pairs in the hot and dense
hadronic medium relative to the vacuum, i.e. an increase of
$\gamma_s$ in Eq. (1). Such modifications might be driven by an
enhanced string tension $\kappa$ in the medium \cite{Soff99} or a relative
decrease of $m_s^2-m_q^2$ with density and temperature.
Alternatively, also the formation of a QGP in the initial phase
might lead to enhanced strangeness production. According to the
authors point of view such phenomena cannot be excluded presently,
however, there is also no strong evidence in favor of them.

\section{UrQMD versus HSD}

Though both transport approaches -- HSD and UrQMD -- give qualitatively
and quantitatively similar results for proton, $\pi^\pm, K^\pm$ and
hyperon rapidity distributions, there are quantitative differences that
should be cleared up.  To this aim we concentrate on idealized model
comparisons which are not directly comparable to experimental data.

\begin{figure}[t]
\centerline{\psfig{figure=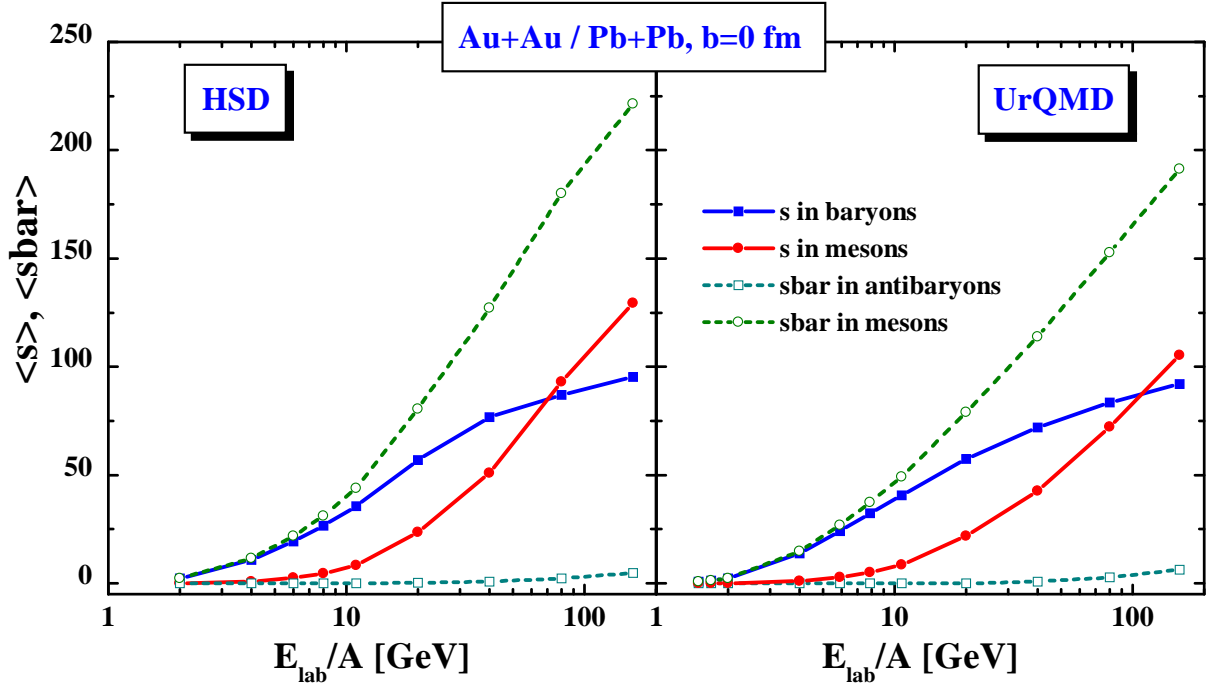,width=16cm}}
\vspace*{5mm}
\caption{The excitation function of $s$ and $\bar{s}$ production from
central $Au+Au$ (AGS) or $Pb+Pb$ (SPS) collisions at impact parameter
$b=0$ for HSD (l.h.s.) and UrQMD (r.h.s.) as appearing in mesons or
baryons (antibaryons).
The solid (dashed) lines with full (open) squares and circles indicate
the number of $s$-quarks ($\bar s$-quarks) in baryons (antibaryons) and
mesons, respectively.}
\label{Fig9}
\end{figure}

The excitation functions for $s$- and $\bar{s}$-quark production in
$b=0$ collisions of $Au+Au$ (AGS) or $Pb+Pb$ (SPS) nuclei (HSD: l.h.s.;
UrQMD: r.h.s.) are shown in Fig.\ \ref{Fig9} which finally end up in
mesons or baryons (antibaryons). The solid (dashed) lines with full
(open) squares and circles indicate the number of $s$-quarks ($\bar
s$-quarks) in baryons (antibaryons) and mesons, respectively.  Both
models give qualitatively and quantitatively similar results (as
already found from the previous comparisons) showing that at $\sim$ 80
A$\cdot$GeV the same amount of $s$-quarks end up in mesons and baryons,
whereas baryons are preferred at lower energies.  The $\bar{s}$-quarks
in antibaryons are almost negligible in this energy range. We mention
here that strangeness conservation is exactly fulfilled in both
transport models such that at all energies the number of $s$-quarks is
identical to the number of $\bar{s}$-quarks.

\begin{figure}[t]
\centerline{\psfig{figure=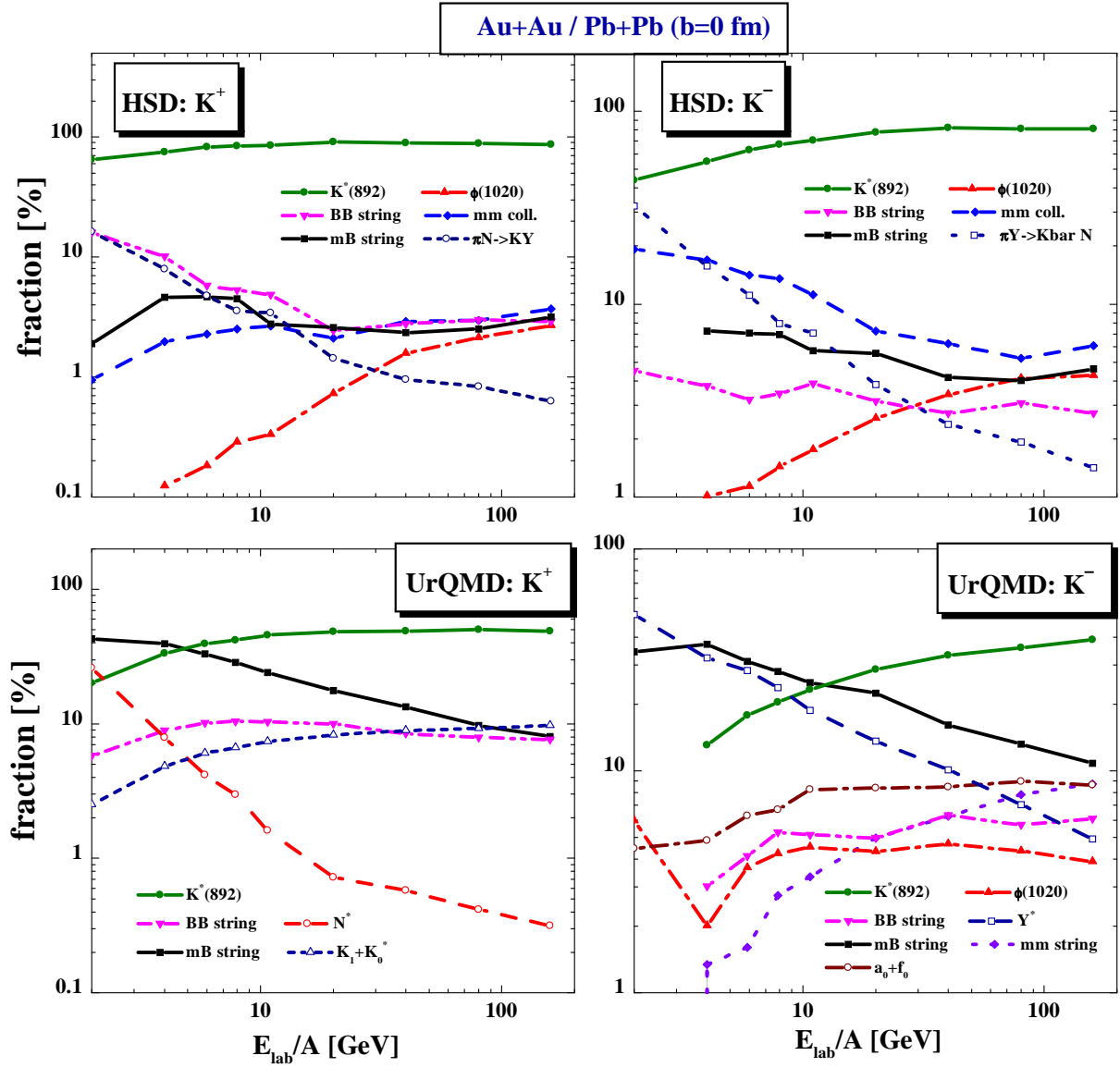,width=16cm}}
\vspace*{5mm}
\caption{The excitation function of the channel decomposition for
$K^+$ and $K^-$ from HSD (upper part) and UrQMD (lower part) from
central $Au+Au$ (AGS) or $Pb+Pb$ (SPS) collisions  at impact parameter
$b=0$.}
\label{Fig10}
\end{figure}

The channel decomposition (fraction in \%) for the finally
observed $K^+$ (left column) and $K^-$ (right column)  from HSD
(upper part) and UrQMD (lower part) are shown in Fig.\ \ref{Fig10}
as a function of bombarding energy for $Au+Au$ (AGS) or $Pb+Pb$
(SPS) collisions at impact parameter $b=0$.  Note, that only the
contributions of the dominant channels are shown in Fig.\
\ref{Fig10}.  For a correct interpretation of Fig.\ \ref{Fig10}
one has to keep in mind, that at high energies initially $s, \bar
s$ quarks are produced in primary nucleon-nucleon collisions and
later on in meson-baryon interactions via string excitation and
decay.  However, subsequently the strange particles (produced
initially) participate in chemical reactions with flavor
exchanges. As a result only a few percent of the 'primary'
kaons/antikaons remain unaffected by secondary inelastic
interactions (cf.~the lines denoted as '$BB$ string' in
Fig.~\ref{Fig10}).  Most of the final $K^+$ and $K^-$ mesons --
above 10 GeV -- finally stem from $K^{*\pm}(892)$ decays (lines
'$K^*(892)$ decay') which are either produced directly in string
decays or by pion-kaon resonant scattering. As seen from the lower
part of Fig.\ \ref{Fig10}, in UrQMD a large fraction of final
$K^+, K^-$ stem directly (without $K^*$ production and decay) from
meson-baryon collisions (line '$mB$ string'). These are realized
in UrQMD via an excitation of a single string that furtheron
decays isotropically, which is reminiscent of the resonance
mechanism. The same mechanism is also used for high energy
meson-meson collisions (line '$mm$ string').  In HSD this channel
is treated differently, i.e.\ via $K^*(892)$ resonance production,
and thus contributes to the '$K^*(892)$' channel.  Note also, that
in UrQMD -- in the channel denoted as '$mB$ string' -- the
kaon/antikaon-baryon collisions are counted, too, whereas they are
not counted here in HSD.  In both models only a few percent of the
final $K^+$ and $K^-$ appear from the $\phi(1020)$ meson decays
(lines '$\phi$ decay'). We mention that only a small fraction of
the $\phi$-decays can be reconstructed from $K^+ K^-$ invariant
mass spectra.

The conceptual differences in the treatment of strangeness production
in both transport approaches are more pronounced at low energies.
UrQMD implements the full resonance dynamics assuming the vacuum
resonance properties: At low energies the baryon and meson resonances
are excited in baryon-baryon ($BB$), meson-baryon ($mB$) or meson-meson
($mm$) collisions (below the string thresholds) and decay (after time
$1/\Gamma$)  to $K^+, K^-$ (cf.  the line '$N^*$' in Fig. \ref{Fig10}
denoting the fraction of the final $K^+, K^-$ from strange baryon
resonance decays:  $N^*(1650)$, $N^*(1710)$, $N^*(1720)$, $N^*(1990)$
and line '$Y^*$' for the higher hyperon resonances as well as the lines
'$K_1+K_0^*$ and $a_0+f_0$ indicating the fraction from the decay of
meson resonances:  $K_1(1270)$, $K_0^*(1430)$ and  $a_0(980)$,
$f_0(980)$, respectively).  In HSD these high strange baryon and meson
resonances are not produced and propagated explicitly (as indicated in
Section 2).  Instead -- in low energy $BB, mB$ or $mm$ collisions --
strangeness is directly produced with respect to the corresponding
transition rates and 2, 3 or 4-body kinematics for the final states
(cf. lines denoted as '$\pi N\to KY$', '$\pi Y\to \bar K N$' or '$mm$
coll.' in the upper part of Fig.\ \ref{Fig10}).  This comparison
demonstrates that the individual channels are treated quite differently
though the final results are very similar.

Thus, as seen from Fig.\ \ref{Fig10}, only a small fraction of
kaons/antikaons (less than 10\% of the kaons and less than 6\% of the
antikaons) from energetic initial baryon-baryon collisions (cf. lines
denoted as $BB$ string) survives the hadronic rescattering phase during
the expansion of the fireball without reinteraction.  Most of the final
strange particles emerge after rescattering --- shifting $s$ quarks from
mesons to baryons and vice versa --- thus providing a very
distorted picture of the initial strangeness production mechanism and
the elementary degrees of freedom involved. Consequently, as pointed
out in Ref.~\cite{WBS_K02}, the $K^\pm$ and $\Lambda$ spectra do not
allow for stringent conclusions on the initial phase of high energy
density. On the other hand, these frequent flavor exchange reactions
may be viewed as the reason why statistical models (employing chemical
equilibrium) seem to work reasonably well.

\subsection{$pp$ reactions}

\begin{figure}[t]
\centerline{\psfig{figure=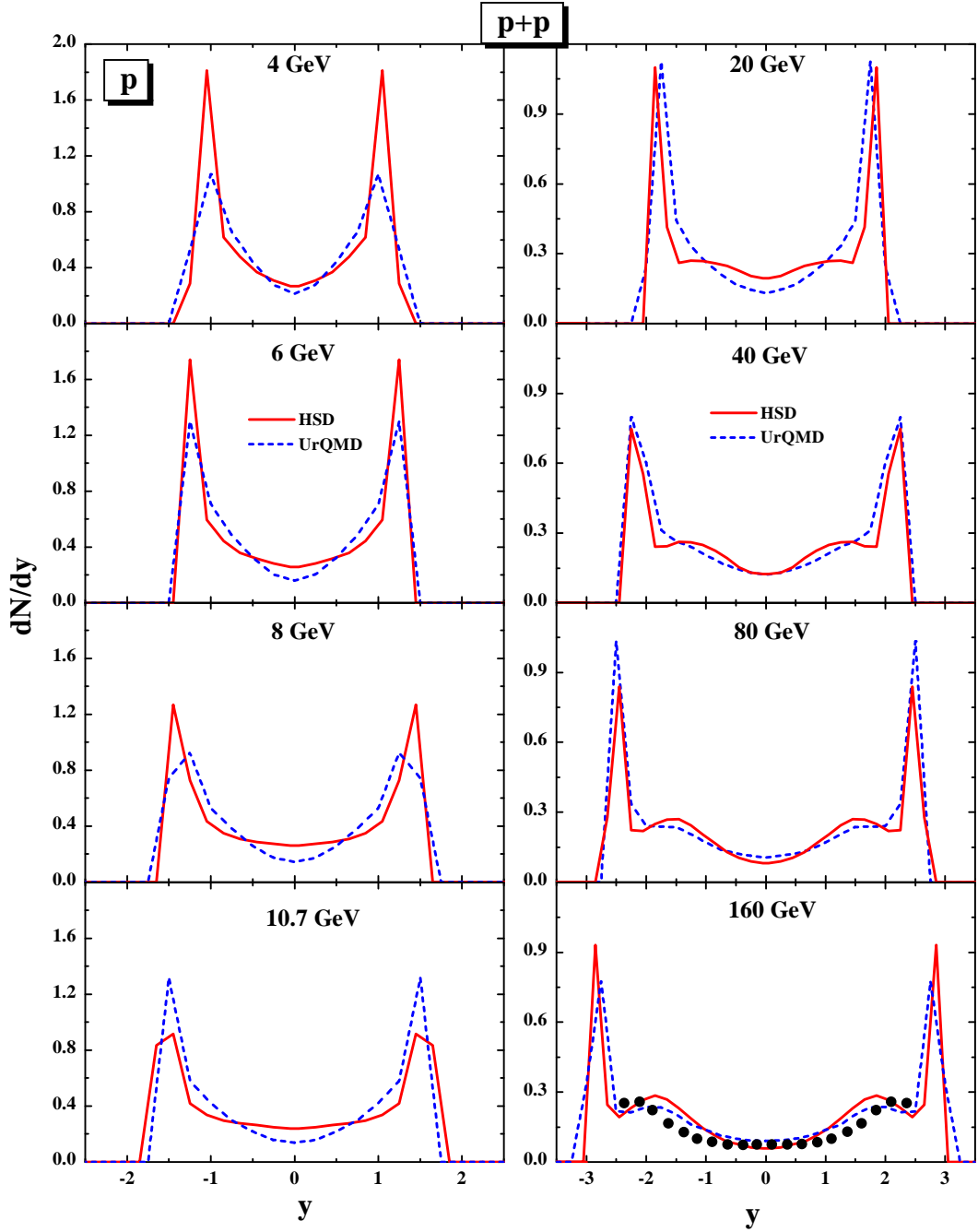,width=14cm}}
\vspace*{5mm}
\caption{The rapidity distribution of protons from $pp$ collisions
at 4--160 GeV as calculated within HSD (solid lines) and UrQMD
(dashed lines).  The experimental data at 160 GeV are taken from Ref.
\protect\cite{NA49pNew}.}
\label{Fig11}
\end{figure}

In order to get some idea about the differences between both
transport approaches we go back to the description
of the elementary channels like $pp$ or $\pi^-p$ in vacuum. In
this respect we show in Fig.\ \ref{Fig11} the proton rapidity
distributions for $pp$ collisions from HSD (solid lines) and UrQMD
(dashed lines) between 4 and 160 GeV laboratory energy. This provides
information on the different string excitation and fragmentation
schemes.  The experimental data at 160 GeV are taken from
Ref.~\cite{NA49pNew}.  As seen from Fig.~\ref{Fig11}, however, the
differences between the two string fragmentation schemes are only
minor. Where do the differences in baryon stopping --- shown
in Fig.~\ref{Fig1} --- come from?

\begin{figure}[t]
\centerline{\psfig{figure=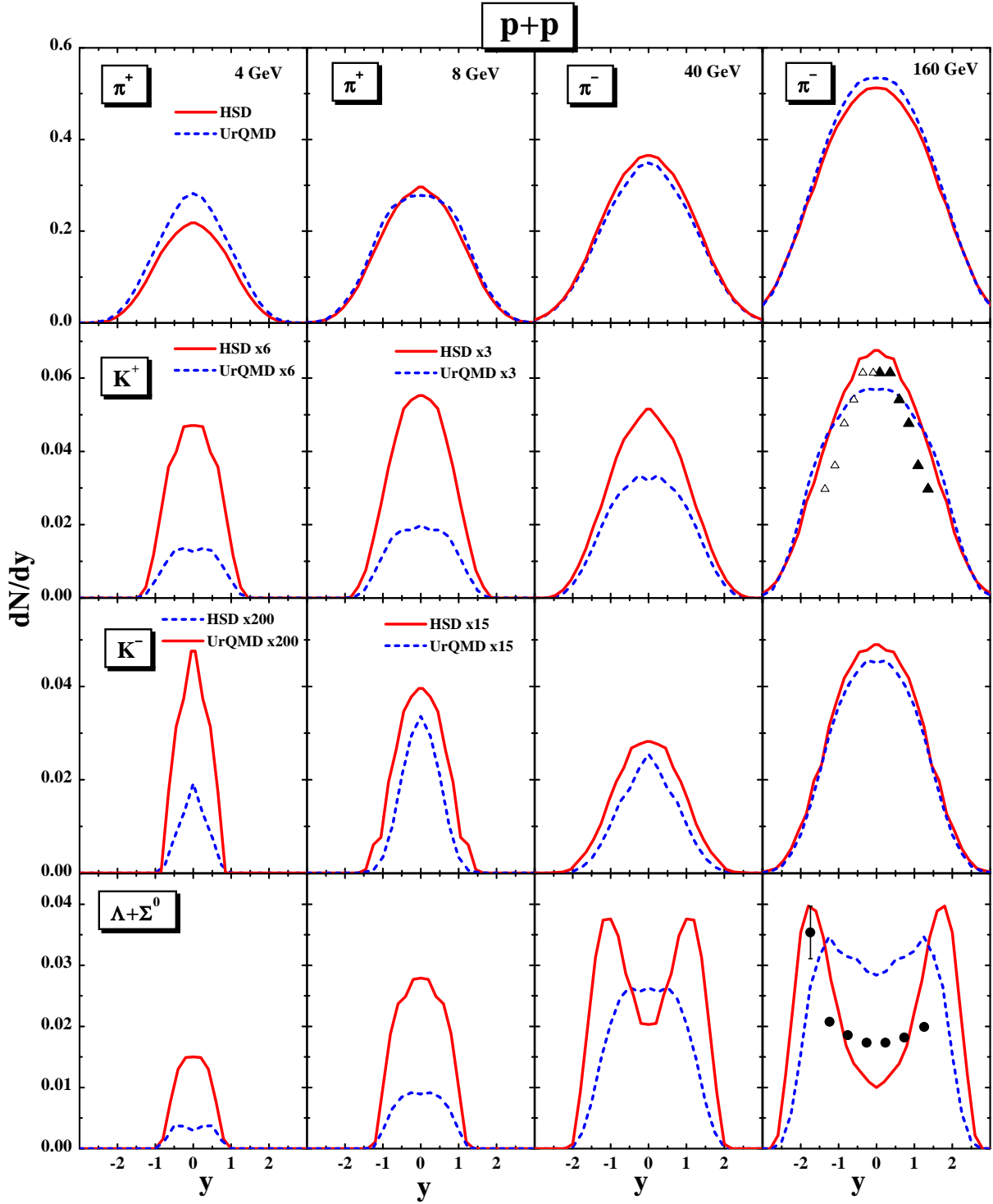,width=16cm}}
\vspace*{5mm}
\caption{The rapidity distribution of $\pi^+, K^+, K^-$ and
$\Lambda+\Sigma^0$'s from $pp$ collisions at 4--160 GeV as calculated
within HSD (solid lines) and UrQMD (dashed lines).
The $K^+$ rapidity distributions from HSD and UrQMD at 4 and 8 GeV
are scaled by  factors of 6 and 3  while the $K^-$ rapidity distributions
are scaled by  factors of 200 and 15, respectively. The experimental data for $K^+$'s
and $\Lambda+\Sigma^0$'s at 160 GeV
are taken from Refs. \protect\cite{Hoehne99} and \protect\cite{NA49ppLam}.}
\label{Fig12}
\end{figure}

The differential rapidity distributions for $\pi^\pm, K^\pm$ and
$\Lambda+\Sigma^0$'s from $pp$ collisions, however, show
substantial differences as demonstrated in Fig.~\ref{Fig12}. The
experimental data for $K^+$'s and $\Lambda+\Sigma^0$'s at 160 GeV
are taken from Refs.~\cite{Hoehne99} and \cite{NA49ppLam},
respectively. Though the $\pi$ rapidity distributions are roughly
comparable in both models, there is a trend for UrQMD to give
slightly more pions with decreasing bombarding energy than HSD,
whereas in heavy-ion collisions the trend is opposite --- HSD
gives more pions at low energies than UrQMD, whereas UrQMD gives
more pions at high energies\footnote{We will attribute this to the
influence of secondary $mB$ scatterings, where more mesons are
produced within UrQMD.} (cf. Figs.~\ref{Fig2}, \ref{Fig3}).  For
$K^-$ mesons the results of both models are comparable at high
energies, however, they deviate closer to the $K^-$ production
threshold.  For $K^+$ and $\Lambda+\Sigma^0$ both models differ
substantially, too. Here HSD gives more  $K^+$ at low energies
whereas the UrQMD rapidity distribution is broader at 160 GeV. The
$\Lambda$ yield from $pp$ collisions is also higher from HSD --
due to strangeness conservation -- and shows distinct peaks in the
rapidity distribution closer to target and projectile rapidities
at high energies, whereas the UrQMD rapidity distributions for
$\Lambda$'s are narrower and almost peaked at midrapidity.
Experimental data \cite{NA49ppLam} -- available at 160 GeV -- show
a minimum at midrapidity giving no strict preference for one of
the string fragmentation schemes.

Thus strange quarks are produced more at midrapidity in UrQMD, both for
mesons and baryons,  whereas in HSD  $s$-quarks are concentrated in
mesons at midrapidity and in baryons at larger rapidities. These
differences in the elementary differential rapidity spectra explain
also the different rapidity distributions from central nucleus-nucleus
collisions in Figs.~\ref{Fig2} and \ref{Fig3} to a large extent.
Presently it is not clear -- due to the lack of corresponding
experimental data -- which fragmentation scheme is 'more realistic'. On
the other hand, this comparison sheds some light on the 'systematic'
uncertainties in present relativistic transport approaches. These
'systematic' uncertainties have to be kept in mind when attempting to
draw conclusions from nucleus-nucleus collisions in comparison to
experimental data.

\begin{figure}[t]
\centerline{\psfig{figure=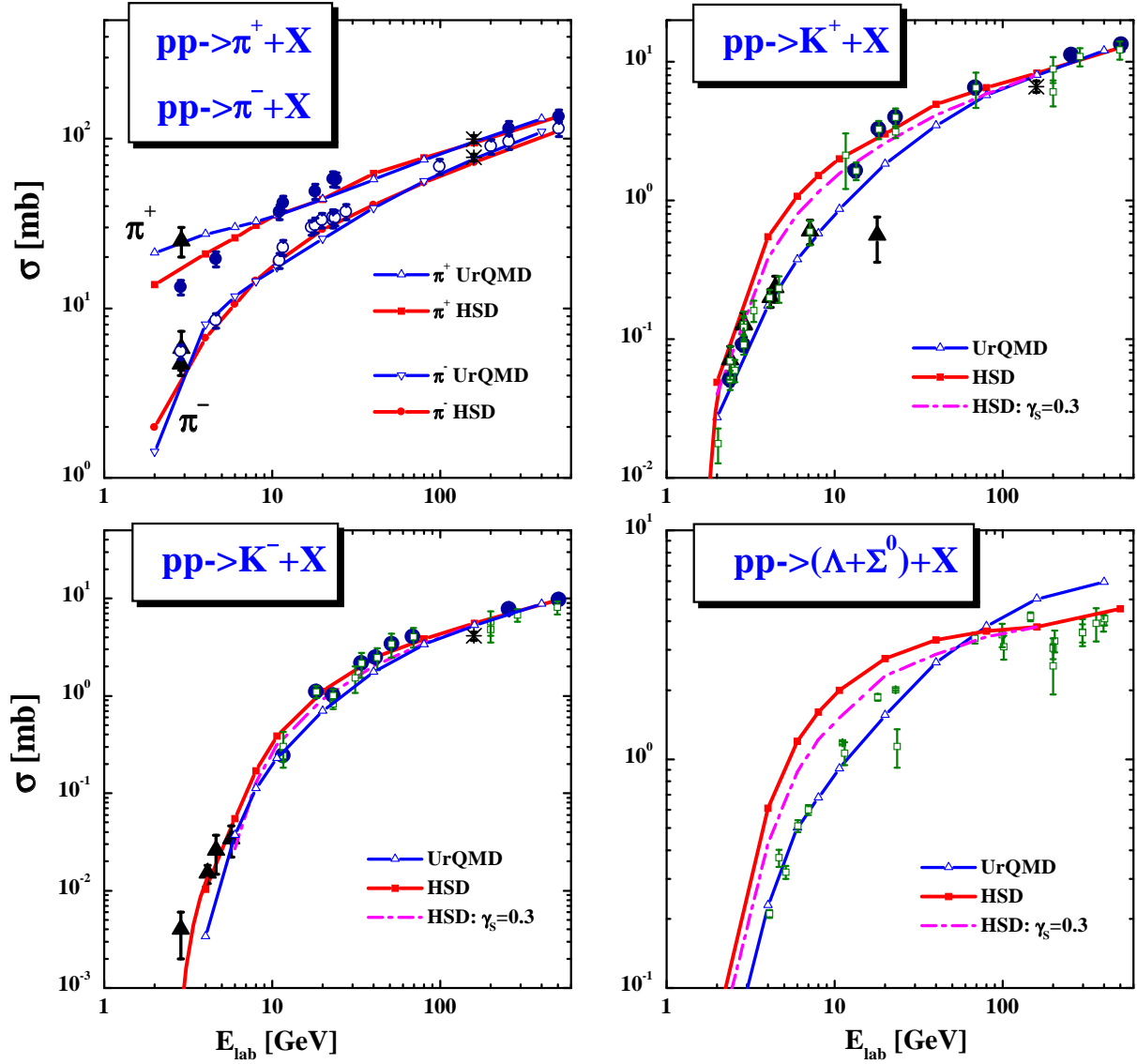,width=16cm}}
\vspace*{5mm}
\caption{The inclusive $\pi^+, \pi^-, K^+, K^-$ and $\Lambda+\Sigma^0$
production cross sections from $pp$ collisions versus kinetic energy
$E_{lab}$. The solid lines with open triangles show the UrQMD results,
the solid line with full squares indicate the HSD results with
$\gamma_s$ defined by Eq. (\protect\ref{supfHSD}), whereas the
dot-dashed lines correspond to the $\gamma_s=0.3$.
The exp. data are taken from Refs. \protect\cite{LB} (full triangles),
\protect\cite{Antin73} (full and open circles),
\protect\cite{Marek_pp} (open squares) and \protect\cite{NA49pp} (stars).}
\label{Fig14}
\end{figure}

Before closing this Subsection we confront both transport models
with the available data on the production cross sections of pions
and strange hadrons from $pp$ collisions.  In Fig. \ref{Fig14} we
show the inclusive $\pi^+, \pi^-, K^+, K^-$ and $\Lambda+\Sigma^0$
production cross sections from $pp$ collisions versus the kinetic
energy $E_{lab}$.  The solid lines with open triangles show the
UrQMD results, the solid line with full squares indicate the HSD
results with the strangeness suppression factor $\gamma_s$ defined
by Eq. (\ref{supfHSD}), whereas the dot-dashed lines correspond to
$\gamma_s=0.3$.  The experimental data are taken from Refs.
\cite{LB} (full triangles), \cite{Antin73} (full and open
circles), \cite{Marek_pp} (open squares) and \cite{NA49pp}
(stars). The pion cross sections are quite well described by both
models; UrQMD gives more $\pi^+$ than HSD at low energy in line
with the data point from \cite{LB}, whereas HSD follows more
closely the data from Ref.  \cite{Antin73}. The inclusive antikaon
cross section is well reproduced by both approaches. As already
demonstrated in Fig.~\ref{Fig12}  HSD gives more $K^+$ and neutral
strange hyperons than UrQMD below $E_{lab} \approx$ 80 GeV. The
neutral hyperon yield from UrQMD (for $E_{lab}\leq$ 80 GeV) is
more in line with the data, whereas the $K^+$ yield is slightly
underestimated from 10 -- 80 GeV. In contrast HSD seems to better
reproduce the $K^+$ cross sections but to overestimate the
$\Lambda + \Sigma^0$ yields in $pp$ reactions at lower energies.

The differences in these 'in-put' cross sections are quite sizeable,
however, one has to keep in mind that only a single isospin channel is
probed in Fig. \ref{Fig14} in comparison to data, whereas in
nucleus-nucleus collisions essentially isospin averaged cross sections
are of relevance. In fact, both transport models differ in the isospin
dependent cross sections for $NN$ collisions, whereas isospin averaged
particle yields are more similar. We recall again that strangeness
conservation holds explicitly for both transport models with respect to
all reactions employed.

\subsection{$\pi^- p$ reactions}

\begin{figure}[t]
\centerline{\psfig{figure=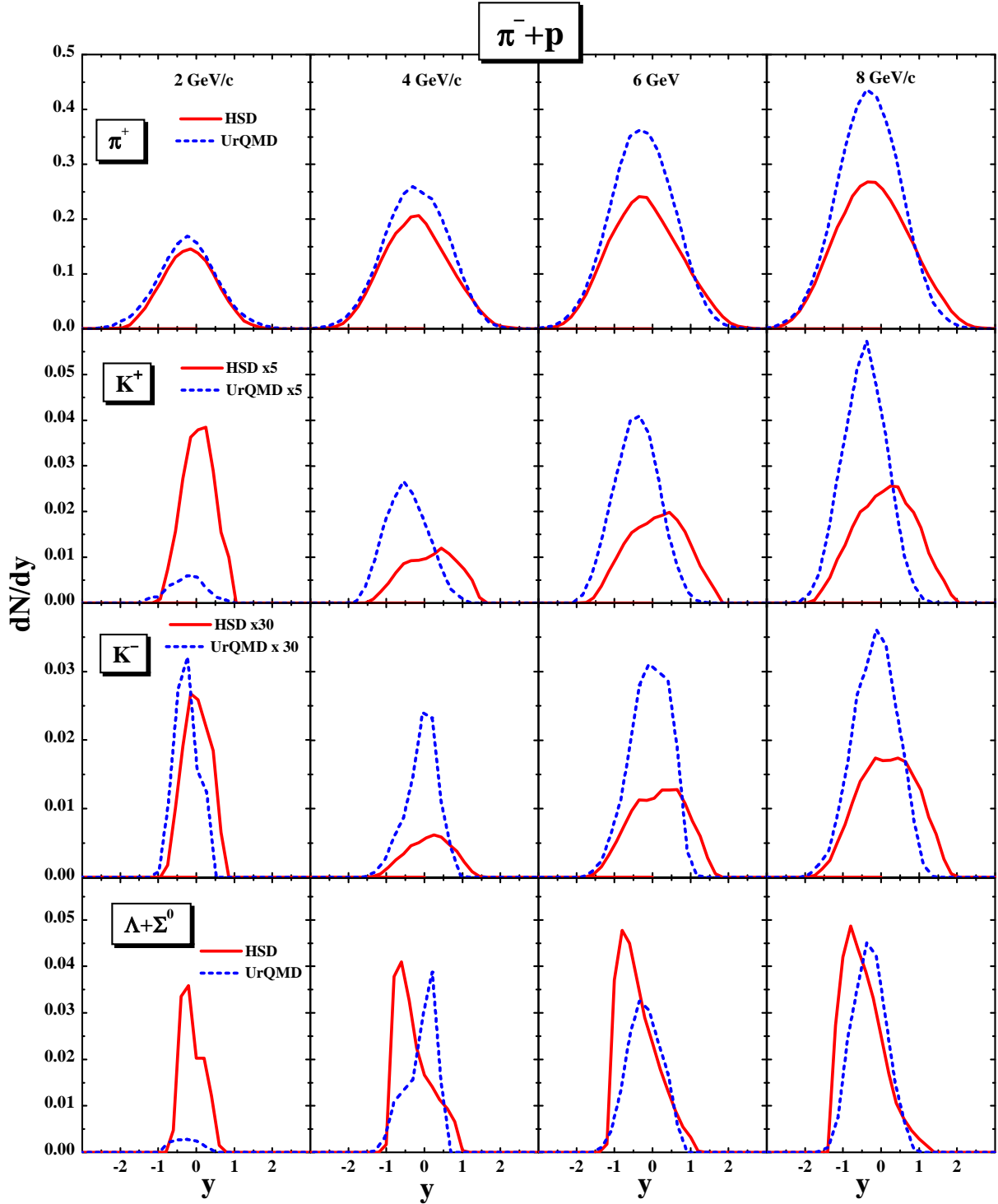,width=16cm}}
\vspace*{5mm}
\caption{The rapidity distribution of $\pi^+, K^+, K^-$ and
$\Lambda+\Sigma^0$'s from $\pi^- p$ collisions at 2--8 GeV/c as
calculated within HSD (solid lines) and UrQMD (dashed lines). Note,
that the $K^+$ and $K^-$ rapidity distributions from HSD and UrQMD  at
2 GeV/c are scaled by factors of 5 and 30, respectively. }
\label{Fig15}
\end{figure}

We now turn to the elementary pion-nucleon collisions that play a
substantial role in secondary meson-baryon collisions. The
differential $\pi^+, K^\pm$ and $\Lambda (+\Sigma^0)$ rapidity
distributions from $\pi^-p$ reactions from 2--8 GeV$/c$ laboratory
momentum are shown in Fig.~\ref{Fig15} for UrQMD (dashed lines)
and HSD (solid lines). Also here experimental data are not
available for a comparison. Though the total and elastic $\pi^- p$
cross sections are very similar in both models and in line with
experimental data (cf. \cite{Jgeiss,UrQMD1}), the explicit
rapidity distributions for various final states differ by up to a
factor of 2-3. This holds true also for the isospin dependent
cross sections (e.g. $K^+$ vs. $K^0$) that are not probed in
nucleus-nucleus collisions due to initial isospin averaging. In
general the string model in UrQMD produces substantially more
$\pi^+, K^\pm$ etc. in $\pi + N$ reactions than the LUND model
employed in HSD. Consequently, hadron (including strange meson)
production by secondary meson-baryon collisions is sizeably higher
in UrQMD than in HSD.  This observation clarifies to some extent
the higher $\pi^\pm$ yield  in Fig.~4 from UrQMD at SPS energies
relative to HSD and the experimental data. On the other hand,
strangeness production ($K^+, \Lambda + \Sigma^0$) from $pp$
collisions is much higher in the LUND approach (cf. Fig.  12) than
in the FSM (used in UrQMD) such that one might expect the HSD
approach to give more kaons and hyperons in central
nucleus-nucleus collisions due to a higher initial production.  As
seen from Figs. 5-7 this expectation does not hold true since in
UrQMD the strangeness production from secondary ($mB$) channels is
substantially higher now, which compensates -- relative to HSD --
for the initially lower strangeness production from $NN$
collisions.

In summary, the dominant differences between HSD and UrQMD for
central nucleus-nucleus collisions can be traced back to different
string fragmentation schemes for $BB$ and $mB$ strings that lead
to substantially different hadron distributions in rapidity as
well as isospin. Presently, these string models are not
sufficiently controlled by differential experimental data.
Furthermore, the string models employed are not tailored to
describe the isospin dependence of the elementary cross sections
at lower invariant energies $\sqrt{s}$.

\subsection{$p p$ versus central $A A$ reactions}
\begin{figure}[t]
\centerline{\psfig{figure=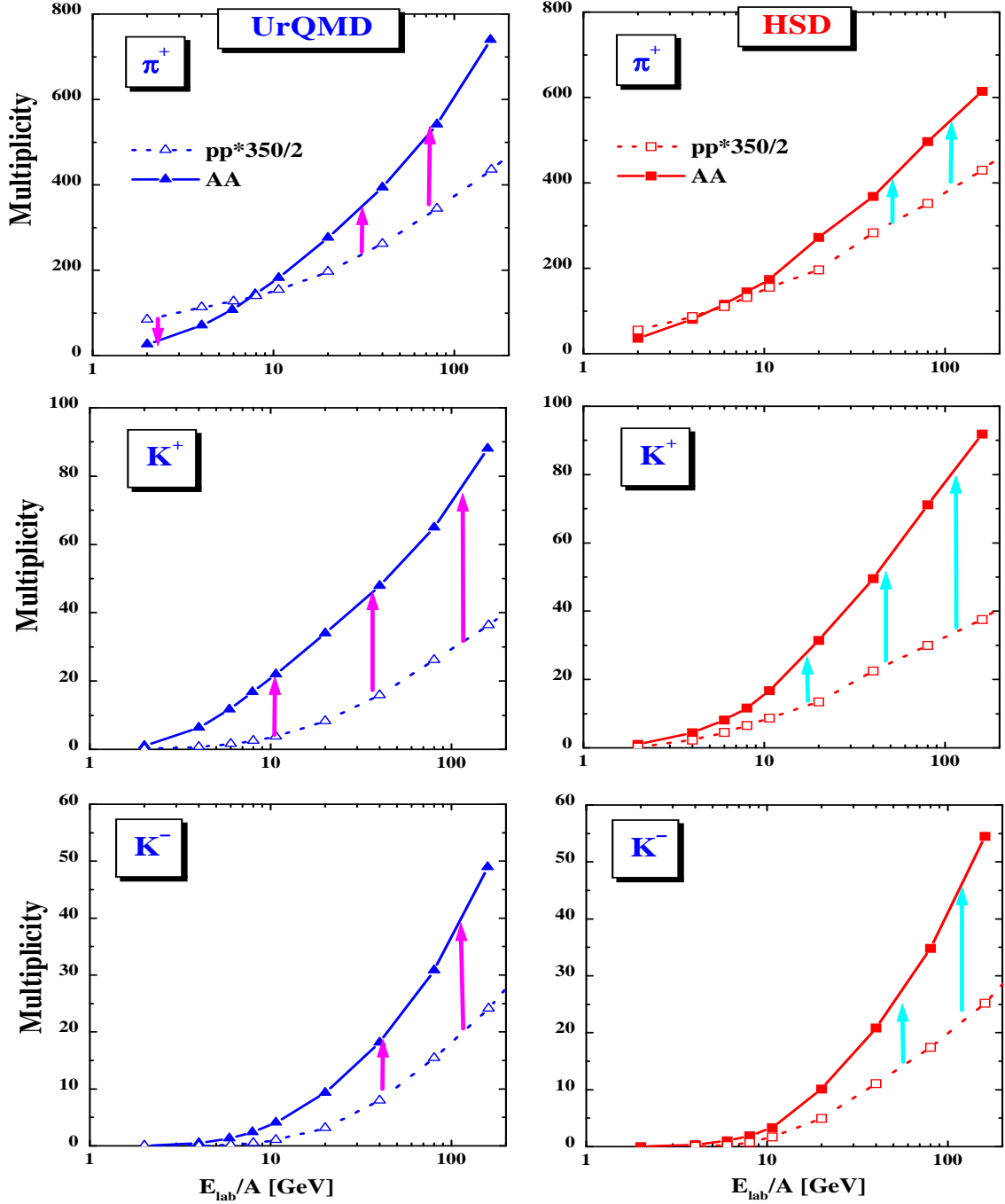,width=15cm,height=18cm}}
\vspace*{5mm} \caption{Total multiplicitis of $\pi^+, K^+$ and
$K^-$ (i.e. $4\pi$ yields) from central $Au + Au$ (at AGS) or $Pb
+ Pb$ (at SPS) collisions in comparison to the total
multiplicities from $pp$ collisions (scaled by a factor 350/2)
versus kinetic energy $E_{lab}$. The solid lines with full
triangles and squares show the UrQMD (l.h.s.) and HSD results
(r.h.s.) for $AA$ collisions, respectively. The dotted lines with
open triangles and squares correspond to the $pp$ multiplicities
calculated within UrQMD (l.h.s.) and HSD (r.h.s.).} \label{Fig16}
\end{figure}

In order to explore the main physics from central $A A$ reactions
it is instructive to have a look at the various particle
multiplicities relative to $pp$ collisions as a function of
bombarding energy. To this aim we show in Fig. \ref{Fig16}  the
total multiplicities of $\pi^+, K^+$ and $K^-$ (i.e. the $4\pi$
yields) from central $Au + Au$ (at AGS) or $Pb + Pb$ (at SPS)
collisions in comparison to the total multiplicities from $pp$
collisions versus the kinetic energy per particle $E_{lab}$. The
solid lines with full triangles and squares show the UrQMD
(l.h.s.) and HSD results (r.h.s.) for $AA$ collisions,
respectively, while the dotted lines with open triangles and
squares correspond to the $pp$ multiplicities calculated within
UrQMD (l.h.s.)  and HSD (r.h.s.). The multiplicities from $pp$
reactions in Fig. \ref{Fig16} have been multiplied by a factor of
350/2 which corresponds to the average number of participants
$A_{part}$ in the heavy-ion reactions for the centrality class
considered divided by the number of participants in the $pp$
reaction. We mention, that the comparison at lower bombarding
energies of 2--4 A$\cdot$GeV has to be taken with some care due to
the different influence of Fermi motion - in case of $AA$
reactions - on the production of pions and $K^\pm$ mesons.

The general trend from both transport approaches is quite similar:
we observe a slight absorption of pions at lower bombarding energy
and a relative enhancement of pion production in heavy-ion
collisions above about 10 A$\cdot$GeV. Kaons and antikaons from
$AA$ collisions are always enhanced in central reactions relative
to scaled $pp$ multiplicities. This enhancement is more pronounced
within UrQMD than in HSD due to the larger cross sections employed
in $\pi N$ secondary reactions as demonstrated in the previous
Subsection.

\section{Summary}

In this work we have performed a systematic analysis of hadron
production in central $Au~+~Au$ or $Pb+Pb$  collisions from SIS to
SPS energies within the HSD (with and without potentials) and
UrQMD transport approaches in comparison to the experimental data
available. We find that both transport approaches --- which are
based on quite different initial ingredients --- roughly give
comparable results for the different $p, \pi^\pm, K^\pm$ and
hyperon distributions in a wide energy regime from 2 -- 160
A$\cdot$GeV. It is remarkable that the cascade mode of HSD (which
operates by default in the potential mode) gives to a large extent
comparable results for strangeness production as the UrQMD
cascade. This observation suggests that --- inspite of the
different elementary 'input' cross sections --- the systems might
reach approximate chemical equilibrium. This is a prerequisite for
an analysis within statistical models
\cite{StatMod,Zschiesche:2002zr,Hahn:mb}.  In fact, the channel
decomposition of strangeness production chains in both models are
quite different (cf. Fig.~\ref{Fig10}) since the degrees of
freedom (hadron resonances and strings)  substantially differ for
collisions at hadron-hadron collision energies around 2-3 GeV in
the region of string thresholds.

We have found that at SPS energies HSD and UrQMD quite well reproduce
the experimental data for $K^-$ and $\Lambda+\Sigma^0$ rapidity
distributions at midrapidity as well as the $4\pi$ yields.  At 20
A$\cdot$GeV both models agree very well among each other for all
hadrons. This provides rather solid predictions for the future GSI
heavy-ion program \cite{GSIprop}.  At AGS energies ($\leq$ 11
A$\cdot$GeV) the $K^+$ yield is slightly overestimated by UrQMD (except
of 10.7 A$\cdot$GeV), whereas HSD underestimats kaon production at the
upper AGS energies (especially with baryon potentials included).  The
$K^-$ and $\Lambda+\Sigma^0$ data are reasonably described by both
models.  We have found also that HSD and UrQMD differ in the pion
multiplicities --- at lower AGS energies the UrQMD model gives slightly
less pions than HSD (with/without potential), but  both models
overpredict the midrapidity data (except UrQMD at 2 A$\cdot$GeV).  At
SPS energies the tendency turns around: UrQMD gives more pions than
HSD, such that HSD is now in a better agreement with the experimental
data.  These differences between the transport approaches  could be
traced back to a large extent to different string fragmentation schemes
which presently are insufficiently controlled by experimental data at
the energies of interest here.

The excitation functions of pions, kaons and antikaons from central 
$Au+Au$ (or $Pb+Pb$) collisions relative to scaled $pp$ reactions from 
the two transport models are very similar: both approaches give an 
absorption of pions at lower bombarding energy and a relative increase 
of pion production for $E_{lab} >$ 10 A$\cdot$GeV.  Kaons and 
antikaons from $AA$ collisions are enhanced in central reactions 
relative to scaled $pp$ collisions at all energies by a factor of 
$\geq$ 2.

We have found that the failure of both models to reproduce the
experimental excitation function for the $K^+/\pi^+$ ratio in central
nucleus-nucleus collisions --- which might suggest the presence of a
different state of hadronic matter in the early phase of these
collisions --- is not primarily due to an underestimation of
strangeness production. Our systematic study in comparison to the most
recent data from the NA49 Collaboration demonstrates that this failure
is mainly due to an inadequate description of pion dynamics. We
attribute this to the fact that the pions in both transport models are
treated as 'free' on-shell particles, i.e.\ with their vacuum
properties and $\delta$-like spectral functions in mass. On the other
hand, lattice QCD as well as effective Lagrangian models indicate an
increase of the pion mass with temperature and density. Furthermore,
the pion spectral function should become broad in the medium due to the
interactions. All these medium modifications have not been included in
the calculations presented in the work. Thus the overestimation of the
pion yields could be a signature for a chiral symmetry restoration
which might occur at the high baryon/meson densities achieved in
relativistic heavy-ion collisions. Including the medium effects for
pions and all strange particles simultaneously in a consistent way in
an 'off-shell transport approach' \cite{Juchem} could provide a more
conclusive interpretation of the experimental data. This, however,
requires a precise knowledge about the momentum and density dependence
of the hadron self-energies in a wide energy regime and full off-shell
transition matrix elements \cite{Juchem}.  Such a program is clearly
beyond the scope of our present study.

Another problem of the transport approaches used here is that detailed
balance is not implemented for $n\leftrightarrow m$ transitions with
$n,m > 2$ \cite{Bravina}. Thus multi-particle collisions might change
the dynamical picture accordingly and lead to 'shorter' chemical
equilibration times \cite{Cass02_antip,Rapp:2000gy,Greiner:2000tu}. In
fact, the importance of $3\leftrightarrow 2$ transitions has been
demonstrated in the extended HSD transport approach in Ref.
\cite{Cass02_antip} for antibaryon reproduction by meson fusion for $A
+ A$ collisions at the AGS and SPS. In order to achieve a more
conclusive answer from transport studies multiparticle interactions
will have to be included in future generations of transport codes.

What to conclude from the detailed comparisons presented in this work?
Coming back to the question raised in the introduction about common
failures in comparison to related experimental data, we can  quote an
insufficient accuracy in the description of the pion degrees of freedom
by both transport models. Does this provide a signal for 'new physics'
in view of a QGP?  The answer of the authors to this question with
respect to the experimental observables studied is: no! As discussed
above, the 'systematic uncertainties' in the 'on-shell' transport
approaches are within the range of the deviations seen in comparison to
the data or even larger. Furthermore, the question raised in the title
of this paper -- anything strange with strangeness? -- also has to be
answered with 'most likely not'!

\section*{Acknowledgement}

The authors acknowledge inspiring discussions with J.~Aichelin,
C.~Greiner, C.~M.~Ko and K.~Redlich. Furthermore, they are indepted to
M.~Ga\'zdzicki, T.~Kollegger, A.~Mischke and M.~van Leeuwen for
providing the experimental data of the NA49 Collaboration in numerical
form.


\clearpage

\clearpage

\clearpage

\clearpage

\clearpage

\clearpage

\clearpage

\clearpage

\clearpage

\clearpage

\clearpage

\clearpage

\clearpage

\clearpage

\clearpage

\end{document}